\documentclass[manuscript]{aastex631}
\usepackage{amsmath}
\usepackage{multirow}

\newcommand{\Msol}[1]{$#1 \, M_\odot$}

\graphicspath{{./}{figures/}}

\received{2022 October 13}
\revised{2023 January 11}
\accepted{2023 January 12}

\submitjournal{MNRAS}

\shorttitle{Hydrogen-rich MWDs}
\shortauthors{Hardy et al.}

\begin{document}

\title{Spectrophotometric analysis of magnetic white dwarf I: Hydrogen-rich compositions}

\author{François Hardy}
\affiliation{Département de Physique, Université de Montréal, Montréal, Québec H3C 3J7, Canada}

\author{Patrick Dufour}
\affiliation{Département de Physique, Université de Montréal, Montréal, Québec H3C 3J7, Canada}

\author{Stefan Jordan}
\affiliation{Astronomisches Rechen-Institut am Zentrum für Astronomie, Universität Heidelberg, Germany}

\begin{abstract}
    We present an homogeneous analysis of all DA stars labeled as magnetic in the Montreal White Dwarf Database (MWDD). Our sample is restricted to almost all known magnetic white dwarf showing clear sign of splitting ($B \gtrsim$ 1-2 MG) that have parallax measurements from the second Gaia data release, photometric data from diverse surveys and spectroscopic data from SDSS or archival data from the Montreal group.
    We determine the atmospheric parameters (effective temperature, surface gravity, magnetic field strength/geometry) of all objects using state-of-the-art model atmosphere/magnetic synthetic spectra, as well as reclassify many objects that were prematurely labeled as potentially magnetic.
    Finally, we discuss the atmospheric parameters/field properties distribution as well as the implication on our understanding of magnetic white dwarfs origin and evolution.
\end{abstract}



\section{Introduction} \label{s:intro}

The study of magnetism in stars is important to understand its role in the formation and evolution of stars in general (fragmentation of protostellar clouds, angular momentum lost, convection, winds, magnetic activity cycles, etc.).
Since characterizing main-sequence magnetic properties can be rather complicated due to their low field strengths, studying their remnants, mostly white dwarf stars, can provide valuable information about the magnetic characteristics of their progenitors as a whole.
Magnetic white dwarfs (MWDs) that are not in an interacting binary system represent around 13 \% of all white dwarfs.
Magnetism in these former stellar cores have been observed over a wide range of strength varying from $10^3$ to $10^9$ G.
The exact origin of the magnetic field in most white dwarf stars is still the subject of debate \citep{Ferrario2015a, Ferrario2015b, Ferrario2020}: are the fields fossil or the results of some binary interaction, merging events or internally produces as core crystallization begins?
Are there signatures in the observed properties of a MWD that allow us to infer its evolutionary channel?

In the fossil field scenario \citep{Tout2004, Ferrario2015b}, the high magnetic fields observed in white dwarf stars are simply the results of magnetic flux conservation as the star evolves (collapses) to a much smaller radius.
Magnetic main-sequence Ap and Bp stars thus appear as natural candidate as progenitors of magnetic white dwarfs (MWDs).
The correspondence between the expected vs. observed field intensities from this population is a strong argument in favor of this hypothesis.
However, the required birth rate of those main-sequence stars is insufficient by a factor of 2 $-$ 3, to account for the incidence of magnetism in WDs \citep{Wickramasinghe2005}.
Alternatively, high magnetic field could be the results of some sort of binary interaction.
The lack of binary systems involving a MWD and a fully detached (and non-degenerate) companion supports this idea \citep{Liebert2015}, at least for a fraction of the population of MWDs \citep{Tout2008}.
Interestingly, MWDs are found to have an average mass that is much higher than their non-magnetic counterparts \citep[average mass of \Msol{\sim 0.8} vs \Msol{0.6},][]{Ferrario2015a}, suggesting the merger evolution branch to be more likely.
It must be noted, however, that \citet{Bagnulo2021} recently showed, based on the local sample (40 pc) that the higher than average mass for MWDs could probably be a selection effect.
Indeed, according to their findings, high field MWDs of normal mass only appear much later on the cooling sequence (possibly as the result of a dynamo mechanism that is triggered by the start of core crystallization) while the high mass ones are present at all temperatures.
Consequently, the overluminous hot MWDs could be over-represented in magnitude limited surveys, skewing the mass distribution toward higher value.

Studying the incidence of magnetism among white dwarf stars, their distribution in terms of mass, effective temperature, field intensity and geometry should provide valuable hints to help us understand the evolutionary paths leading to MWDs.

The incidence of magnetism among white dwarf stars is still uncertain as volume-limited samples and magnitude-limited sample yield values ranging from 10 to 20\% \citep{Kawka2007, Giammichele2012, Bagnulo2021} to 2-5\% \citep{Liebert2003, Kepler2015} respectively.
However, those numbers certainly suffer, at the lowest field, from the fact that spectropolarimetric measurements have been obtained only for a small fraction of stars \citep[and references therein]{Vennes2018, Kawka2007, Kawka2019, Landstreet2012, Landstreet2015, Bagnulo2022}.
Moreover, the low signal-to-noise spectra for many objects barely allow the line splitting to be resolved \citep[and references therein]{Kepler2015, Kepler2019}.
It is to be noted that the number of known magnetic white dwarfs has increased substantially in the last two decades or so.
Indeed, fewer than 70 objects were known at the turn of the century \citep{Wickramasinghe2000} but thanks mainly to the Sloan Digital Sky Survey \citep[SDSS,][]{York2000} there are allegedly now more than 800 objects that have tentatively been identified as MWD \citep{Gansicke2002, Schmidt2003, Vanlandingham2005, Kulebi2009, Kepler2013, Kepler2015, Ferrario2015a}.
About 75\% of these are isolated stars with estimated magnetic field strength between $10^3$ G to $10^3$ MG, the other 25\% being in interacting binary systems \citep{Ferrario2020}.

However, our knowledge of the bulk properties of MWDs with magnetic fields in the MG regime relies mostly on pioneering work done in the 1980’s and 1990’s \citep[for example][]{Wickramasinghe1979, Jordan1991, Jordan1992, Bergeron1992, Wickramasinghe2000, Schmidt2003}.
The last comprehensive homogeneous analysis of a large hydrogen-rich MWD sample (100+ object), that of \citet{Kulebi2009}, is now more than a decade old.
It must be noted that most studies, including that of \citet{Kulebi2009}, have been carried out assuming $\log{g} = 8$ as no parallax measurements were available for most objects at the time of analysis.
Another reason is that, as mentionned in \citet{Ferrario2020}, a common method for the determination of the surface gravity for the non-magnetic WDs is performed using the broadening of spectral lines due to the Stark effect.
Since no calculations of line profiles taking both the Stark and magnetic broadening into account exists, determination of $\log{g}$ using the spectrum only is not reliable.

Given the large number of MWDs that have been identified since that study, and the mass availability of precise trigonometric parallax measurements for basically every object in the solar neighborhood \citep{Gaia2016, Gaia2018, Gaia2021}, it is now more than necessary to revisit/determine the atmospheric parameters of all known MWD in a homogeneous fashion as such an analysis is bound to provide important clues as to the nature and evolution of magnetic white dwarfs in general.

As a first step, this paper will look at all the objects that have been classified as magnetic hydrogen-rich white dwarfs in the Montreal White Dwarf Database \citep{MWDD}.
In \autoref{s:observations} we present the observations and our sample selection and in \autoref{s:theory} we describe our theoretical framework and the description of our analysis method for the photometric as well as the spectroscopic part of this work.
\autoref{s:results} present the details of our spectrophotometric analysis while \autoref{s:discussion} discusses those results and implications on our understanding of the evolution of MWDs.

\section{Observations} \label{s:observations}

\subsection{Sample Selection} \label{ss:sample}

As a first step, we aim to analyze in a homogeneous fashion all known magnetic hydrogen-rich white dwarfs that have been identified in the literature over the years \citep[][and references therein]{Ferrario2015a, Ferrario2020}.
We also include the many objects that have been flagged as DAH by visual inspection of thousands of SDSS spectra \citep[mostly from][]{Kepler2015, Kepler2016}.
Most of these magnetic white dwarfs have never, to our knowledge, been the subject of a detailed analysis.
Furthermore, we restrict ourselves to objects showing clear signs of magnetic line splitting ($B$ greater than about 2 MG), as the analysis of weakly MWDs requires either much higher resolution and signal-to-noise ratio spectroscopic observations than what is currently available (the bulk of our sample being from SDSS) or spectropolarimetric measurements \citep[such as][and references therein]{Vennes2018, Kawka2019, Bagnulo2021, Bagnulo2022}.
We thus first selected all the objects with the spectral type DA and \verb|Magnetic| flag in the Montreal White Dwarf Database \citep[hereafter MWDD,][194 stars]{MWDD} or with spectral type DAH (510 objects).
Stars that did not clearly show signs of magnetic splitting, or were known to have a magnetic field strength well below 1 MG based on published values in the literature were removed from our sample (see below).
We also removed a few objects that were modeled as a DA+DC unresolved binary \citep{Rolland2015} as the analysis of such systems requires special treatment.
While it is possible that a few genuine high field magnetic DA white dwarfs may have been missed, given that the MWDD includes all the MWD reported in the compilations of \citet{Ferrario2015a, Ferrario2020} as well as the new identifications from \citet{Kepler2015, Kepler2016}, we are confident that our sample contains almost all known hydrogen-rich MWD with fields higher than 2 MG that have been identified in the literature.

This leaves us with a sample of 661 objects.
However, a few of these objects had no optical spectra available to us, leaving us with a final sample of 651 potentially MWDs to examine.
Each of these objects is then analyzed in a homogeneous fashion using the best available spectra (see MWDD), astrometric measurements from Gaia DR2/EDR3 \citep{Gaia2016, Gaia2018, Gaia2021} and photometric measurements (either SDSS \emph{ugriz} \citep{York2000}, Pan-STARRS \emph{grizy} \citep{PanSTARRS} or \emph{JHK} and \emph{BVRI} \citep{Bergeron1997, Bergeron2001}, see \autoref{ss:photo-spectro-data}).

\begin{figure}
    \centering
    \includegraphics[width=0.7\linewidth]{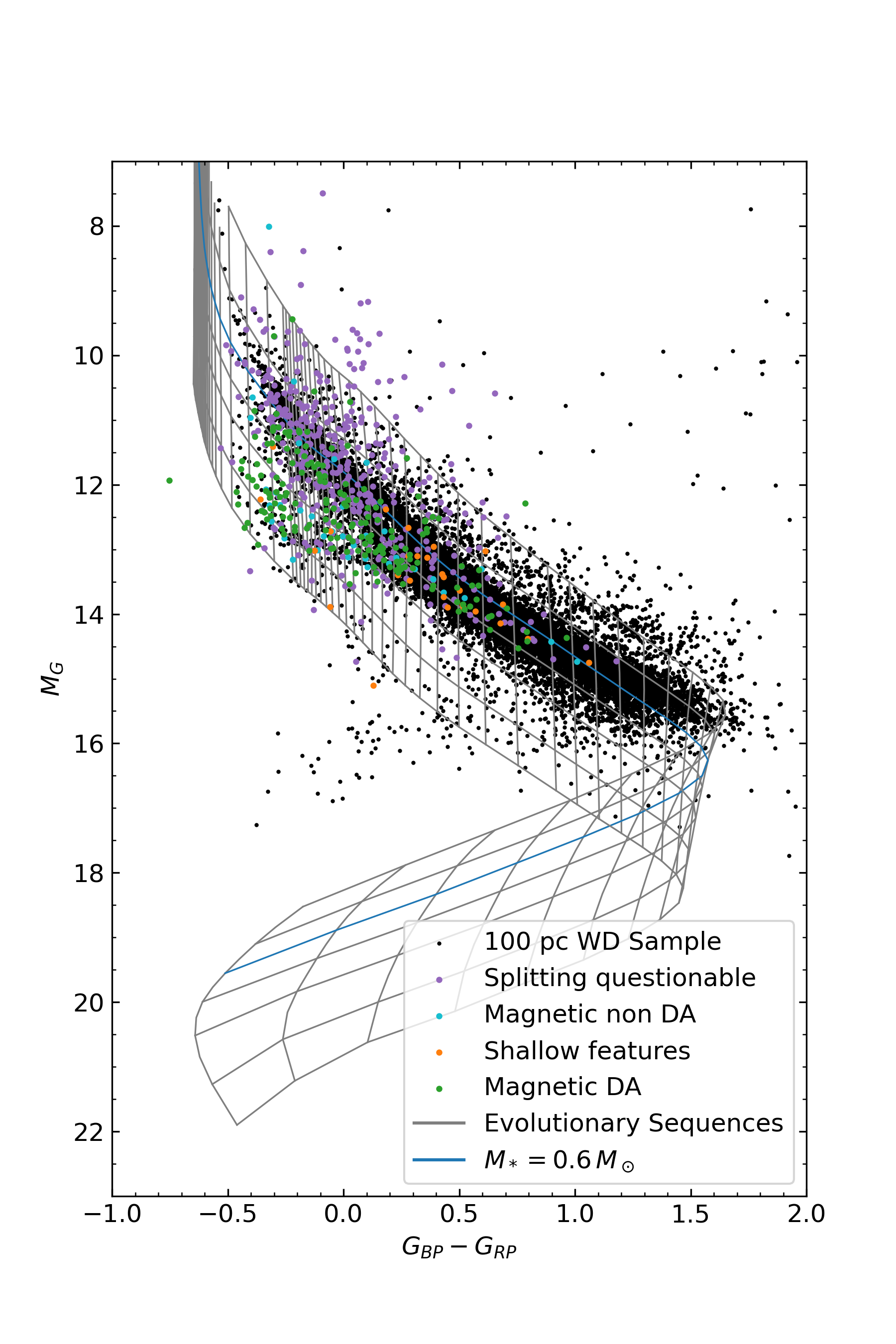}
    \caption{\label{f:color_mag_gaia} Gaia H-R diagram for our sample of magnetic DAs. Black dots are the white dwarfs from the Gaia EDR3 within 100 pc. The various color dots represent our full sample of 651 stars, classified into categories explained in \autoref{s:results}.}
\end{figure}

\autoref{f:color_mag_gaia} shows an observational Hertzsprung-Russell (H-R) diagram for the objects in our sample (colored dots) as well as WDs within 100 pc (black dots).
Evolutionary sequences for non-magnetic DAs are also shown.
The classification of the colored dots is the result of our analysis described in \autoref{s:results}.
The figure shows a clear separation between the MWD DAs and the rejected MWD candidates (probably normal DAs, see below), with MWDs tending to have lower luminosity, smaller radius, and therefore a higher mass than their non-magnetic counterparts.

\subsection{Photometric and Spectroscopic Data} \label{ss:photo-spectro-data}

In order to perform an analysis as homogeneous as possible, we decided to use SDSS \emph{ugriz} point-spread function magnitudes in priority when available.
When \emph{ugriz} data were not available, we relied on Pan-STARRS \emph{grizy} or previously published \emph{JHK} and \emph{BVRI} photometry, in that order of priority.
In the end, out of 651 objects, only 29 (4.5\%) had no \emph{ugriz} data.
Hence we used \emph{grizy} data for 26 stars and \emph{BVRI} photometry for the last 3.

The majority (598 out of 651, 92\%) of the spectra used for this work are collected from SDSS.
The spectra of the 53 objects without SDSS spectroscopy were taken from \citet{Limoges2013, Limoges2015}, \citet{Bergeron1997, Bergeron2001}, \citet{Gianninas2011}, \citet{Giammichele2012}, \citet{Tremblay2020} or archival data obtained by the Montreal group over the years.

\subsection{Note on Naming Convention} \label{ss:names}

Numerous names are often used to design a given white dwarf. In order to avoid confusion, we decided to follow a convention similar to that of \citet{Coutu2019} and use names based on ICRS coordinates at epoch and equinox 2000 (alternative names for any given star can easily be obtained through the search function on the Montreal White Dwarf Database).
Stars are thus referenced with JHHMM$\pm$DDMM, where HHMM corresponds to the right ascension (R.A.) in hours and minute and DDMM to the declination (decl.) in degrees and minutes in sexagesimal notation.
\autoref{t:names} provides the correspondence between the names of all objects in our sample with Gaia source ID, MWDD ID, and the coordinates in decimal format (degrees).

\section{Theoretical Framework} \label{s:theory}

\subsection{Magnetic Synthetic Spectra Calculations}

The presence of a strong magnetic field has a strong effect on both the atomic energy levels, line strengths as well as the radiative transfer as polarization has to be taken into account.

\begin{figure}
    \centering
    \includegraphics[width=0.7\linewidth]{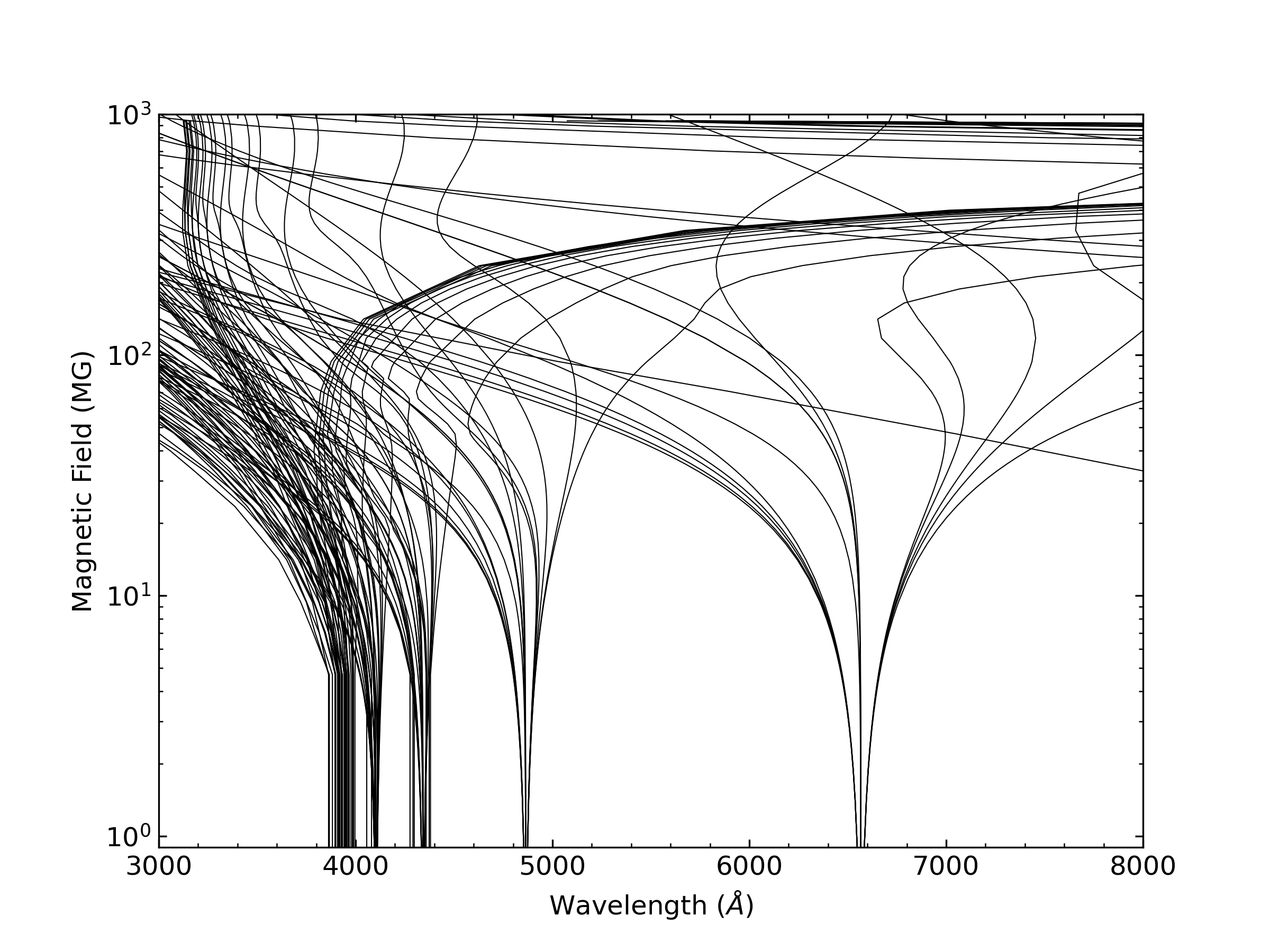}
    \caption{\label{f:balmer} Position of the Balmer lines between 3000 and 8000 \AA{} as a function of magnetic field strength.}
\end{figure}

We rely on data provided by one of us \citep[S.Jordan, similar from][]{Forester1984, Wunner1985, Wunner1989} which provide transitional energies, wavelengths, and various quantum quantities needed to compute oscillator strengths for a wide range of magnetic field intensities from 0 to over 9000 MG.
The data used in this work were never fully published, and are denser than the ones available in the literature.
As can be observed in \autoref{f:balmer}, line positions vary substantially in a non-linear way for fields above about 10 MG.
The spectral features for objects with such strong fields can thus become much more complex than the usual Zeeman triplet found at lower fields.
We implemented an interpolating scheme developed by \citet{Steffen1990} to obtain the component data for arbitrary field intensities, with a linear extrapolation when required.
This extrapolation occurs at over 9000 MG for most components, although some weaker components only have calculations up to 1000 MG.
As most stars in our sample have a dipolar magnetic field $B_p$ of under 100 MG with only a very limited number of stars with $B_p$ over 500 MG, this extrapolation is not expected to be frequent.
We also obtained similar data for helium transitions and implemented them into our routines.
This work assumes only pure hydrogen atmospheres (other compositions will be explored in Hardy et al., in preparation).

An appropriate polarized radiative transfer routine also had to be implemented for a correct description of MWD radiation field.
\citet{Jordan2003} describe four methods to numerically solve the radiative transfer equations, and we decided to use their MATEXP method (Section 2.4 of their work), as it yielded smoother polarization spectra, was more numerically stable for our case, and was one of the fastest in computation time.
While in principle our code is now suitable for the analysis of other Stokes parameters, we restrict our analysis to the Stokes parameter $I$ (the luminous intensity) in this work as no polarimetric data is yet available for the majority of our sample\footnote{It should be noted that this is not the same as \emph{neglecting} the other Stokes parameters, as observational data is not yet widely available.}.

\begin{figure}
    \centering
    \includegraphics[width=0.7\linewidth]{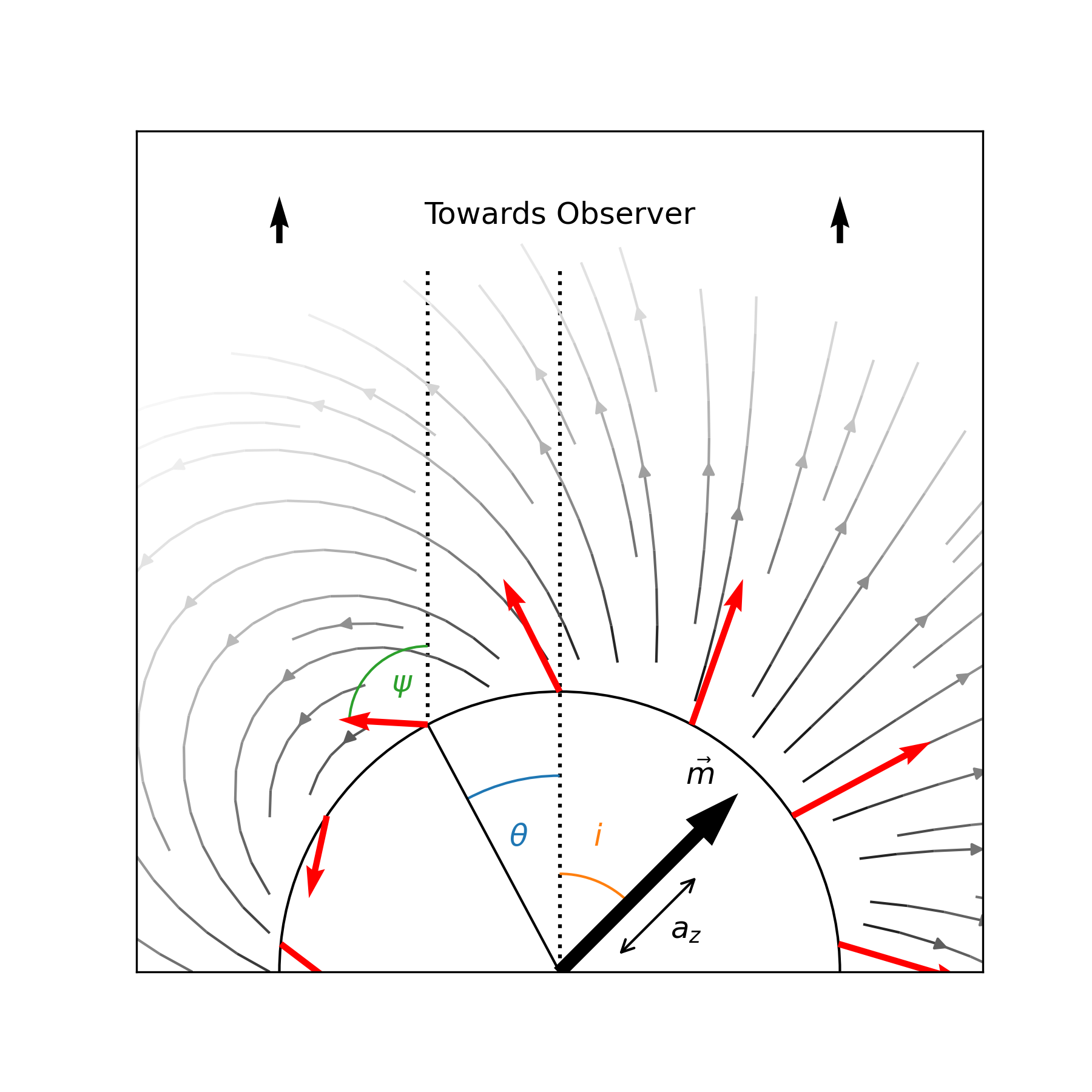}
    \caption{\label{f:dipole_schema}Schema for the various parameters describing our dipole geometry. In red are the magnetic field vectors at the stellar surface, with the angle $\psi$ in green and $\theta$ in blue used to model each surface elements. $\vec{m}$, the inclination angle $i$ and dipole offset $a_z$ describe the geometry of the field.}
\end{figure}

As a first-order approximation, we model the surface magnetic field of the stars with an inclined and offset dipole:

\begin{equation}\label{e:dipole}
    \mathbf{B}=\frac{3\left(\mathbf{m}\cdot \hat{\mathbf{r}}\right)\hat{\mathbf{r}}-\mathbf{m}}{{r}^{3}}
\end{equation}

\noindent where:

\begin{itemize}
    \item[] $\mathbf{B}$ is the magnetic field vector, in MG,
    \item[] $\mathbf{m} = \frac{B}{2} \hat{\mathbf{m}}$ is the dipole moment vector ($B$ is the dipole intensity, in MG, $\hat{\mathbf{m}}$ the unit vector of the dipole moment),
    \item[] $\hat{\mathbf{r}}$ is the unit vector of the position vector $\mathbf{r}$, from the dipole to where the field is measured (here, the surface of the star),
    \item[] $r$ is the absolute value of the position vector $\mathbf{r}$, in stellar radii.
\end{itemize}
We orient the magnetic moment according to the inclination, and apply the dipole offset by shifting the position vector accordingly.
The parameters used to describe the dipolar geometry (\autoref{f:dipole_schema}) are the dipole intensity $B_p$, which is the magnetic field intensity at the pole of a centered dipole, at the stellar surface ($1 \, R_*$ from the origin), the inclination angle $i$ between the dipole moment $\vec{m}$ and the observer's line of sight, and the dipole offset $a_z$ (in stellar radii $R_*$), the distance between the center of the dipole and the center of the star.

We discretize the stellar surface into many surface elements.
The number of surface element has to be large enough to correctly sample the variation of the magnetic field on the surface, but not too large as to slow down the calculation significantly.
We find that using 149 surface elements was optimal as no significant change to the computed synthetic spectra were found in tests using a finer resolution.
Once the specific intensity for each surface element has been calculated, we integrate across the stellar surface to obtain the Eddington flux for any given geometry.

In order to recover the magnetic field geometry for every star in our sample, a significant number of synthetic spectra across a wide range of effective temperatures, surface gravity, and geometries were required.
Pre-generating all the necessary dipolar-geometry intensity spectra would have been much more time consuming (and less expandable in the future) than integrating them on-the-fly using a precomputed grid of spectra with a constant field.
We therefore built a spectrum grid across 5 parameters: effective temperature (5000 K to 30,000 K in 15 steps), surface gravity ($\log{g}$ from 7.0 to 9.5 in 6 steps), field intensity ($10^{-2}$ MG to $10^4$ MG in 159 steps), the angle between the line of sight and the normal of the surface ($\psi$ from 0 to $\pi$ in 6 steps), and the limb darkening ($\mu = \cos{\theta}$ from 0 to 1 in 4 steps).
From this precomputed grid (85 GB of data), a synthetic spectrum with any desired field strength and geometry (not limited to offset dipole) can be reconstructed rapidly.
Only offset dipole geometries are considered in this exploratory work.

\subsection{Photometric Analysis Technique} \label{ss:photo}

The first step in our analysis is to estimate the effective temperature and surface gravity of our objects.
However, the presence of a strong magnetic field can affect significantly the determination of these atmospheric parameters.
As a start, we first obtained effective temperature and surface gravity with models with no magnetic field.
The atmospheric structures and synthetic colors used are those of \citet{Bergeron2019} \citep[same atmospheric models used by][]{GenestBeaulieu2019}.
Note that those models do not include the effect of magnetism on the stellar structure.
In fact, it is not clear at the moment how exactly the structure is affected by the presence of a strong magnetic field and as a consequence, all past studies, to our knowledge, used non-magnetic structure to calculate magnetic synthetic spectra \citep[see][for example]{Kulebi2009}.
We thus employed a similar strategy here but improved on the fitting technique by including explicitly the effect of magnetism when determining atmospheric parameters from photometric data (see below).

We obtain the effective temperature and surface gravity with zero-field models using an evolutionary algorithm \citep[based on][]{Storn1997} to fully explore the parameter space of the grid (this is a test to make sure we recover the non-magnetic parameters obtained with standard technique before proceeding with magnetic synthetic spectra).
We applied de-reddening corrections to the observed magnitudes to take into account the amount of interstellar material on the line of sight \citep{Schlafly2011}.
We then constructed new synthetic grids with various magnetic field strengths to account for the shifting of lines from one photometric band to the other as the field increases.
The atmosphere structures are the same as those used previously but the synthetic spectrum used to derive the synthetic photometric colors include the effect of the magnetic field on the line splitting.
With these photometric grids in hand, we proceeded to obtain a value for the atmospheric parameters for every star in our sample for field strength of 0, 10, 50, 100, 500, 700, and 900 MG.
For each star in our sample, we built an interpolating table for the atmospheric parameters (effective temperature, surface gravity), plus mass and radius, so we can get interpolated values for arbitrary magnetic field intensities that will be obtained in the next step.
While this is done assuming a constant surface magnetic field, this method allows to at least capture the main effect of the presence of strong magnetic fields, that is the shifting of lines in adjacent bands (the exact details of the geometry is a second-order effect in that matter).

\begin{figure}
    \centering
    \includegraphics[width=0.8\linewidth]{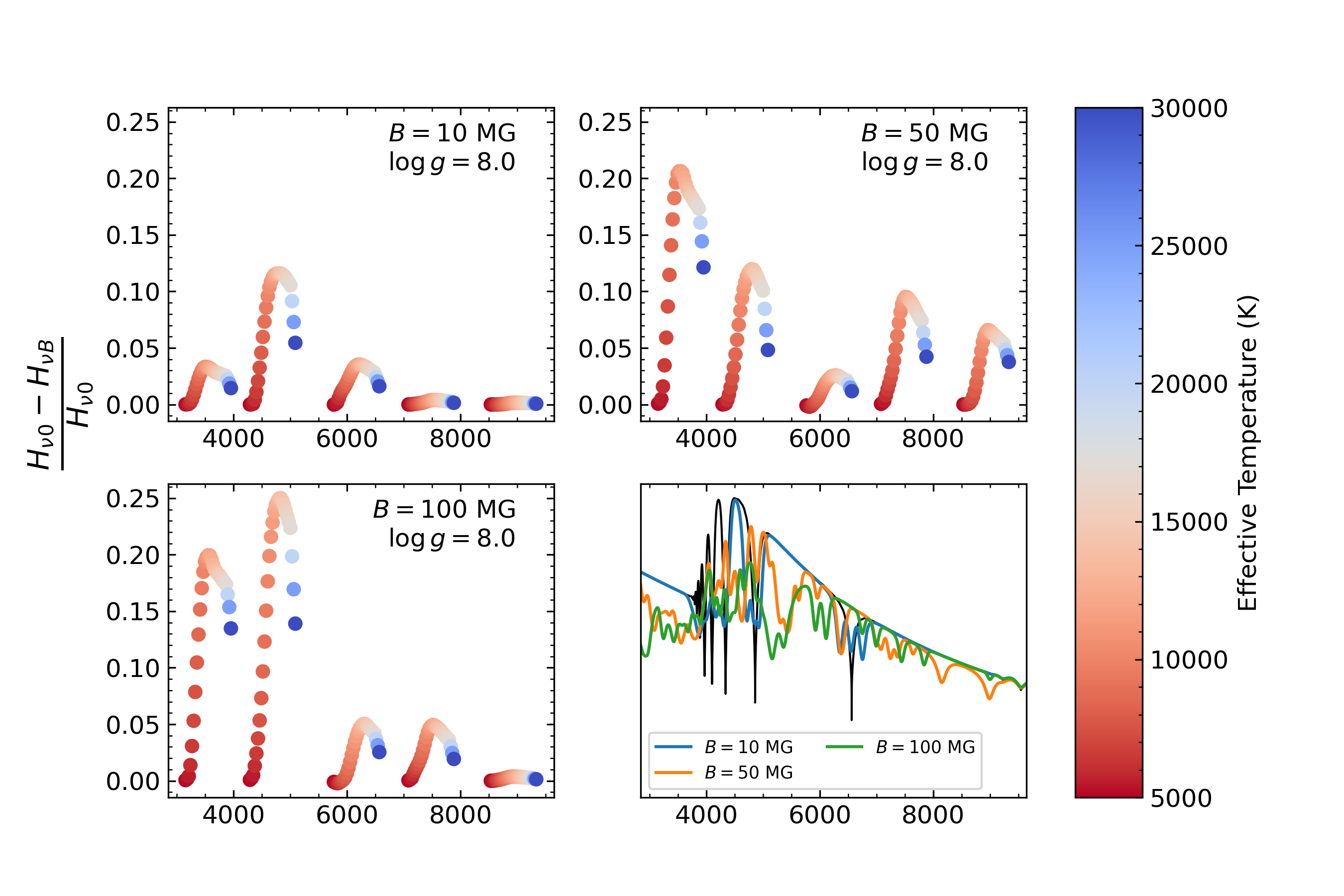}
    \caption{\label{f:magnitudes}Relative differences in magnitude for models with and without magnetic field.}
\end{figure}

\autoref{f:magnitudes} shows how the synthetic \emph{ugriz} colors are affected by the inclusion of magnetic splitting for various field strength and effective temperatures.
In the case of magnetic DA stars, the \emph{u} and \emph{g} bands are the most affected, since the Balmer lines (except H$\alpha$) are near them.
With weak or no magnetic field, there are barely any Balmer lines in the \emph{u} band, but a magnetic field with an intensity above a few MG pushes many components inside the band.
A similar phenomenon happens for the \emph{g} band, which is in the range of the Balmer lines starting from H$\beta$.
Here the many split components affect the \emph{g} band considerably. With increased magnetic field, the \emph{r} band is the least affected as H$\alpha$ components tend to stay inside the band and the \emph{i} band can only have H$\alpha$ components drift inside.
The \emph{z} band is a little different, since some of the Paschen lines are not split in our model, due to a lack of data.

Including the effect of magnetism in the synthetic spectra can thus have a non-negligible effect on the atmospheric parameter determination (effective temperature and surface gravity).
This is the first time, to our knowledge, that this effect is explicitly taken into account for the determination of the atmospheric parameters.
Uncertainties on the fitted values were obtained by fitting the photometry again, but with a least-squares minimization method that also returns the estimated uncertainties.
We used the values obtained with the fitting method described above as starting values for the algorithm, ensuring uncertainties around those values.

\subsection{Spectroscopic Analysis Technique} \label{ss:spectro}

When fitting a given star, the process to obtain a synthetic geometry-aware spectrum is the following:
\begin{itemize}
    \item Provide the dipole geometry parameters: $B_p$, $i$, $a_z$;
    \item Generate an $N$-element discretized surface geometry and compute the average surface field intensity;
    \item Interpolate the effective temperature and surface gravity to that average field value;
          \begin{itemize}
              \item See previous \autoref{ss:photo}.
          \end{itemize}
    \item Use those atmospheric parameters with each surface element to obtain the $N$ spectrum elements by interpolating in the 5-D grid;
    \item Sum the $N$ spectra into the dipolar-geometry-aware spectrum at the desired atmospheric parameters.
\end{itemize}

We use this process in the fitting technique to obtain every spectrum needed for each iteration until convergence.  
Due to the high dimensionality of the problem and the highly degenerate nature of the solutions, standard minimization methods using the gradient descent are ill suited for this situation, since many local minima are found with different dipolar geometry.
The usual steepest-descent algorithm would get stuck on the first minimum it would encounter based on an initial guess value.
Making this initial guess value would be different for each star and is not optimal.
This is why the optimization algorithm used relies on a stochastic nature and allows searching for a wide area inside the parameter space.
It is part of the \verb|scipy| Python package, which is an implementation of the algorithm by \citet{Storn1997}.

We tested our method by generating non-magnetic and magnetic synthetic stars with various magnetic field intensities and geometries and fed it to our fitting code.
Our algorithm was able to retrieve the geometry for both the magnetic and the non-magnetic stars with satisfactory precision.
For the special case of non-magnetic synthetic stars, we still used the algorithm to get a dipolar geometry, but the dipolar intensity from the solution is minimal and the other parameters (angle and offset) are less constrained and are of no physical value.

Typical uncertainties were estimated by changing the dipole parameters ($B_p$, $i$, $a_z$) until the resulting synthetic spectrum diverged significantly from the observations, for a typical magnetic star in our sample.
We observed that the relative uncertainties for the different parameters were very different, meaning the solutions were more sensitive to an increase of magnetic field intensity ($B_p$) than the other geometry parameters.
We found a relative uncertainty of 5\% for the $B_p$ parameter, 20\% for the inclination $i$, and 50\% for the dipole offset $a_z$.

In summary, the fitting technique for magnetic spectra begins with the creation of a random set of ($B_p$, $i$, $a_z$) parameters.
For each individual in this population, we go through the process described above to obtain a spectrum, and calculate the $\chi^2$ value against the observations.

\section{Spectrophotometric Analysis} \label{s:results}

Using the fitting technique described in \autoref{s:theory}, we fitted all 651 objects in our sample in a homogeneous way assuming single DA synthetic spectra with a dipolar geometry.
Each solution was then carefully inspected by eye to judge the quality of the fit. As a result of this laborious exercise, we could classify all our objects in broadly 5 categories (see \autoref{t:class_count}): i) stars that are probably not magnetic (400) ii) stars well fitted with a dipole geometry (140) iii) stars that are clearly hydrogen-rich magnetic white dwarfs but that are moderately well modeled with a dipolar geometry (45) iv) stars that have hydrogen features too shallow (28) and finally v) stars that we were simply unable to fit using the current theoretical framework (38). We now discuss each category in turn.

\begin{deluxetable}{cccc}
    \tabletypesize{\footnotesize}
    \tablecolumns{4}
    \tablewidth{0pt}
    \tablecaption{\label{t:class_count} Numbers of stars in our different categories}
    \tablehead{
        \colhead{Category} & \colhead{Subcategory} & \multicolumn{2}{c}{Count}
    }
    \startdata
    \multirow{2}{*}{Magnetic DA}     & Good dipole fit  & 140 & \multirow{2}{*}{185} \\
    & Bad dipole fit   & 45  &                      \\
    \hline
    \multirow{3}{*}{Dismissed stars} & Magnetic non DA / Unable to fit  & 38  & \multirow{3}{*}{466} \\
    & Shallow features & 28  &                      \\
    & Splitting questionable       & 400 &                      \\
    \hline
    \textbf{Total}                   &                  & \multicolumn{2}{c}{651}
    \enddata
\end{deluxetable}


The first category contains more than half of our original sample (400 stars).
Most of these object had been classified as DAH or DAH: from visual classification of low signal-to-noise ratio spectra from various data release catalog of spectroscopically confirmed white dwarf stars from the Sloan Digital Sky Survey \citep{Kleinman2004,Kleinman2013,Kepler2015,Kepler2019}.
While visual inspection indeed seems to show possible shifted components near the core of some Balmer lines (thus the DAH classification), it quickly became clear that those classifications did not hold up against detailed comparison with magnetic synthetic spectra.
For example, it was possible in a few cases to force a magnetic field strength/geometry to fit some glitches near the core of one Balmer line only to find that such configuration lead to predicted components that are not observed in the other lines.
When considering all the lines simultaneously, our algorithm was not able to find a magnetic solution that was statistically better than a non-magnetic fit. In the end, although it is quite possible that some of these stars will be found to be genuinely magnetic white dwarfs when higher signal-to-noise or spectropolarimetric observations become available, we believe that classifying these objects as DAH based on these low quality SDSS spectra was perhaps a bit premature.
We report in \autoref{t:B_lt_1MB} the photometric atmospheric parameters (effective temperature and $\log{g}$) assuming $B = 0$ and reassign them with the spectral type DA until proof to the contrary.


\startlongtable
\begin{deluxetable}{lccc}
    \tablecaption{\label{t:B_lt_1MB} Stars with no compelling evidence of magnetism}
    \tablehead{
        \colhead{J name} & \colhead{$T_{\rm{eff}}$ (K)} & \colhead{$\log{g}$} & \colhead{Mass ($M_\odot$)}
    }
    \startdata
    J0002$-$4310 &   8979 (328) &     8.49 (0.3) &    0.912 (0.38) \\
    J0003$-$0153 &  15408 (638) &    8.49 (0.17) &    0.918 (0.18) \\
    J0005+1934 &   5367 (248) &    7.57 (0.11) &   0.360 (0.074) \\
    J0007+2407 &   7653 (660) &    7.27 (0.25) &    0.268 (0.14) \\
    J0010+2429 &    6247 (46) &    7.74 (0.76) &   0.448 (0.072) \\
    J0012+5025 &    6354 (27) &   8.23 (0.031) &   0.735 (0.041) \\
    J0016+2350 &  12296 (378) &    9.09 (0.61) &    1.233 (0.56) \\
    J0040+0702 &  16927 (454) &    8.61 (0.17) &    0.997 (0.22) \\
\enddata
\end{deluxetable}

\begin{figure}
    \centering
    \includegraphics[width=\linewidth]{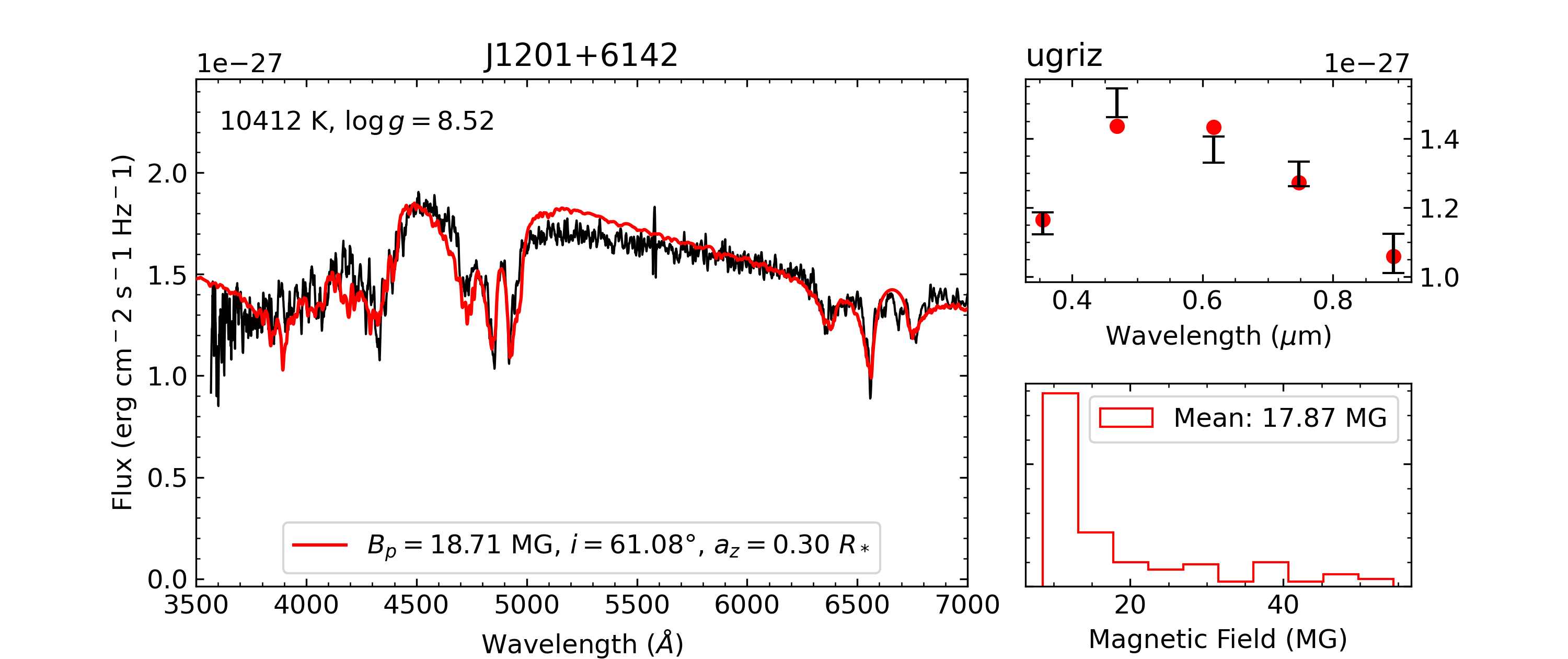}
    \caption{\label{f:good_fit} An example of a star that is successfully reproduced by a dipolar field geometry. The left panel is our best fit to the Balmer lines. Upper right panel represents the best fit to the $ugriz$ photometry while the bottom right panel shows the distribution of magnetic field elements at the visible surface of the star for our best fit solution.}
\end{figure}

\begin{figure}
    \centering
    \includegraphics[width=\linewidth]{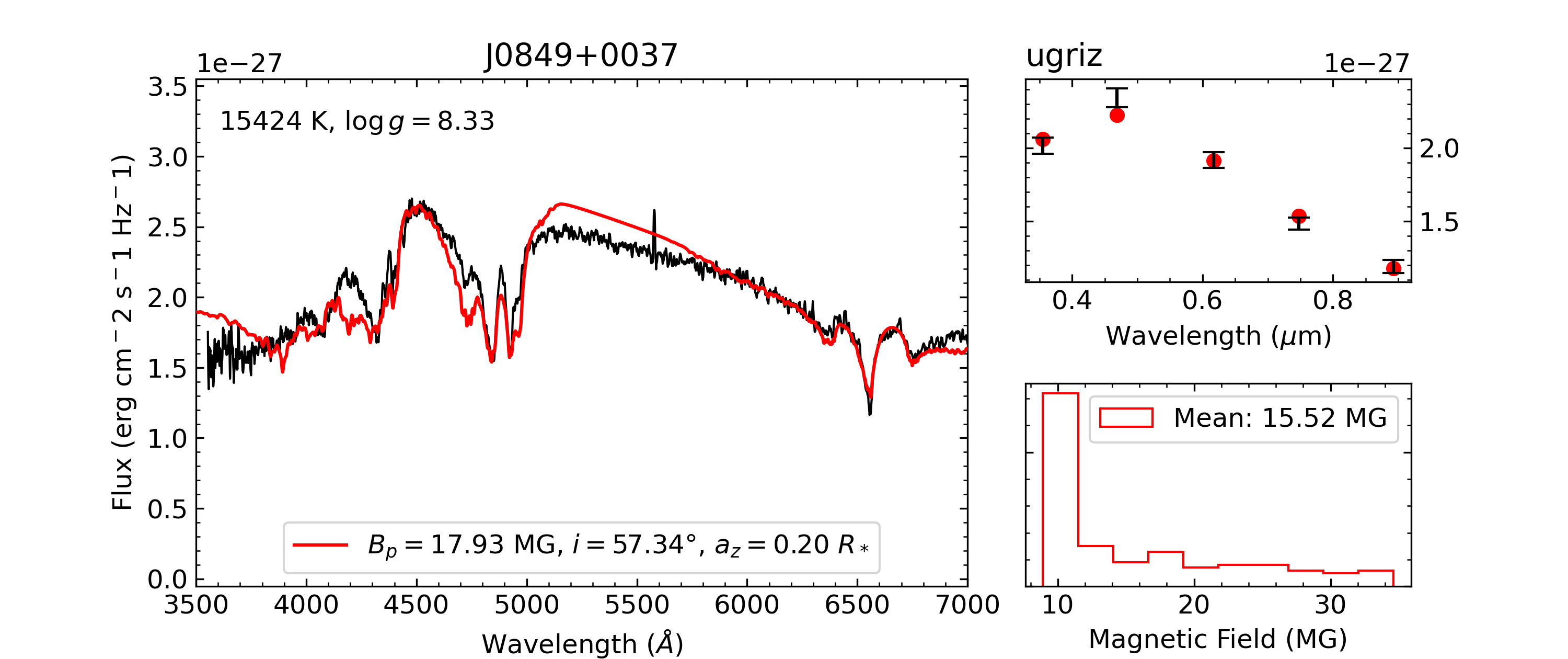}
    \caption{\label{f:ish_fit} An example for a star where a shifted dipole provides a bad fit.}
\end{figure}

The next category comprises 140 objects for which an off-centered dipole model did a pretty decent job at reproducing the various magnetically shifted components.
\autoref{f:good_fit} shows an example of such stars.
While in some cases some minor discrepancies could be observed between the model flux and observations, the overall quality of the fit is deemed sufficient to label them as \emph{Good dipole fit}.
On the other hand, the quality of the dipole fit was clearly inferior for the third category (45 objects) that we call \emph{Bad dipole fit}.
\autoref{f:ish_fit} shows an example of such bad fits.
That so many objects cannot be well reproduced with a dipolar geometry should not be surprising.
For one, a shifted dipolar field geometry is probably only a first-order approximation for a much more complex field structure and we should not expect all objects to be easily modeled this way.
Second, it is well known that many MWDs show rapid photometric variations on timescales of minutes to hours \citep[see for example][and references therein]{Williams2022, Kilic2021}.
A good example is that of G183$-$35 \citep{Kilic2019}, an unusual white dwarf that shows an H$_\alpha$ line split into five components, instead of the usual three components seen in strongly magnetic white dwarfs (we also find five components for J1328+5908 and J1430+2811). Time-resolved spectroscopy of this object seem to support the idea that we are witnessing rotational modulations of a complex magnetic field structure.
As our sample is composed mostly of SDSS spectra that were integrated on a period of time that span an unknown fraction of the total period, it is no surprise that many objects are not well reproduced with a simple dipole. These objects, listed in \autoref{t:results_bad}, represent good candidates to study the complexity of magnetic field structure through time-resolved spectroscopy.

Nevertheless, although the overall quality of the fits for these stars are clearly inferior in terms of reproducing the exact shape of all the Balmer components, the resemblance between the best fit models and the observations suggests that the determined mean surface fields are probably in the right ball park.
As a consequence, we can be confident that the determined effective temperatures and surface gravities/masses for these objets are also closer to reality than those determined from non-magnetic models. All our fits are presented in Appendix \hyperref[f:all_fits]{A1}.

\begin{figure}
    \centering
    \includegraphics[width=\linewidth]{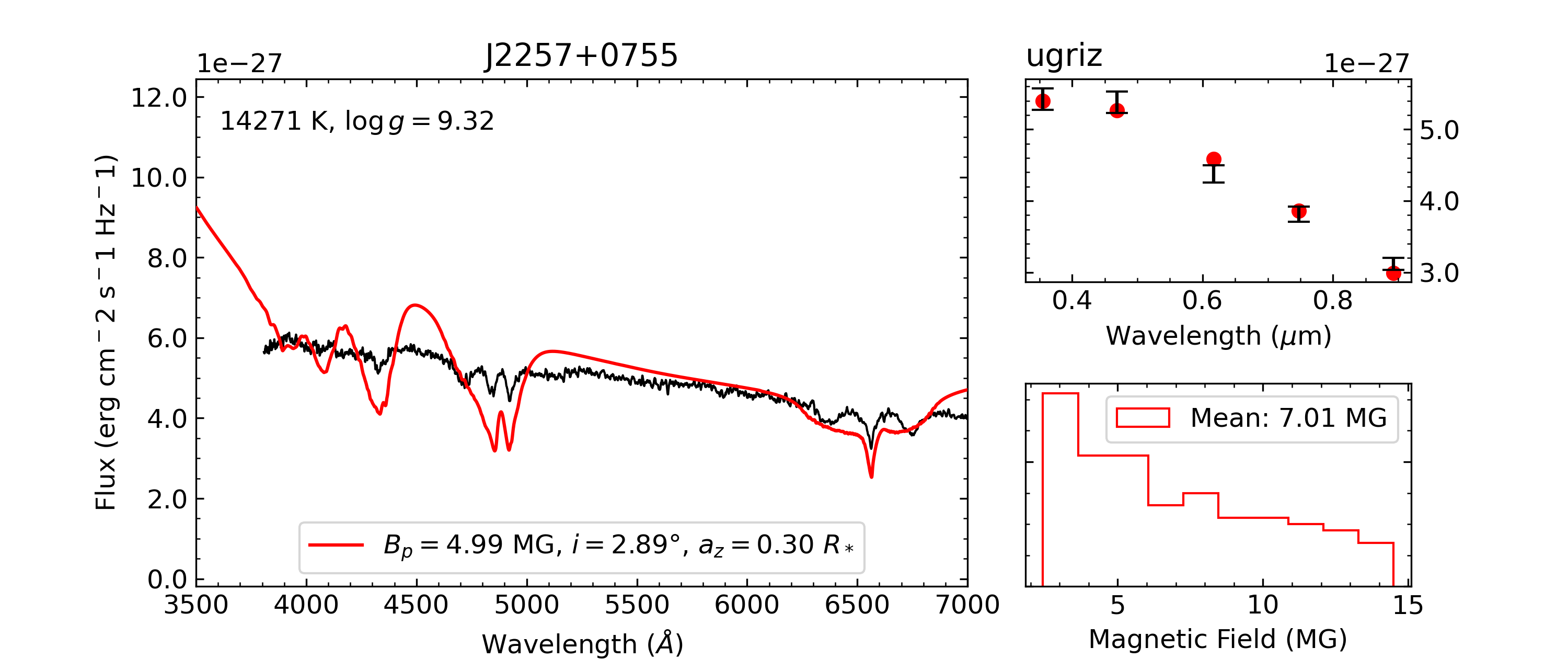}
    \caption{\label{f:shallow_features} J2257+0755 (EGGR 156), an example of a star whose observed features are too shallow, according to our models. Due to the fitting algorithm trying to minimise the numerical difference between the synthetic spectrum and observations with shallow features, the magnetic structure here is not representative of reality.}
\end{figure}

Another interesting category found within our sample contains objects with hydrogen features that are very shallow compared to what magnetic model spectra predict for the fitted atmospheric parameters and amount of splitting observed (28 objects, \autoref{t:shallow}).
\autoref{f:shallow_features} shows an example of such shallow features.
It must be noted that no combination of inclination, offset or field strength can reproduce those spectra.
It is also highly unlikely that a more complex multipole expansion could alleviate the issue as the depth of the central component of each line has only a weak dependence on the geometry.
In the case of J2257+0755 (EGGR156), \citet{Kulebi2009} managed to obtain a good fit by increasing the effective temperature to 40,000 K.
This solution, however, is clearly incompatible with the shape of the spectral energy distribution ($ugriz$) which requires an effective temperature in the vicinity of 14,000 K.
Lowering the effective temperature below $\sim 10,000$ K also reduces the depth of the absorption features, but again, at the cost of not reproducing the $ugriz$ photometry anymore.
Similar objects with shallow features have been modeled has DA+DC unresolved binary in \citet{Rolland2015} but in the case of J2257+0755, the observed photometric modulations rules out a close binary scenario and the star is rather interpreted as the rapidly rotating remnant of a double degenerate merger \citep{Williams2022}.
If so, rapid rotation may also produce an unusual surface composition that may explain the shallow lines found in such stars.
For example, an attractive solution to the absorption lines depth problem is that such objects have a mixed H-He composition.
We experimented with mixed atmosphere and indeed found that it was possible to reduce significantly the depth of the lines in that manner.
We calculated models with increasingly more helium, until the depth of hydrogen lines were well reproduced, and found that a 12,000K helium-dominated atmosphere with $\frac{N\left({\rm He}\right) }{N\left({\rm H}\right)} = 10^4$ can reproduce H$\alpha$ while not predicting helium features that are not observed.
Unfortunately, such models badly reproduce H$\beta$.
It is possible that the atmosphere is not homogeneous, with patches of different composition and temperature, but exploring the full range of possibilities is beyond the scope of this paper.
Again, time-resolved spectroscopy may eventually help us uncover the real nature of such objects.
We will thus postpone the analysis of these stars to future studies and exclude them from our discussion below.

\startlongtable
\begin{deluxetable}{lllll}
    \tablecaption{\label{t:shallow} Stars with shallow features}
    \tablehead{
        \colhead{} & \colhead{} & \colhead{J name} & \colhead{} & \colhead{}
    }
    \startdata
    J0343$-$0641 & J0906+0807 & J0944$-$0018 & J1057+0411 & J1132+2809 \\
    J1137+5740 & J1201+0847 & J1248+4104 & J1251+5432 & J1314+1732 \\
    J1315+0937 & J1328+5908 & J1332+0117 & J1344+3754 & J1418+3123 \\
    J1430+2811 & J1516+2803 & J1639+1036 & J1816+2454 & J2115+0400 \\
    J2131+0659 & J2138+1123 & J2151+5917 & J2211+1136 & J2223+2319 \\
    J2257+0755 & J2259$-$0828 & J2346$-$1023 &  &  \\
    \enddata
\end{deluxetable}

\begin{figure}
    \centering
    \includegraphics[width=\linewidth]{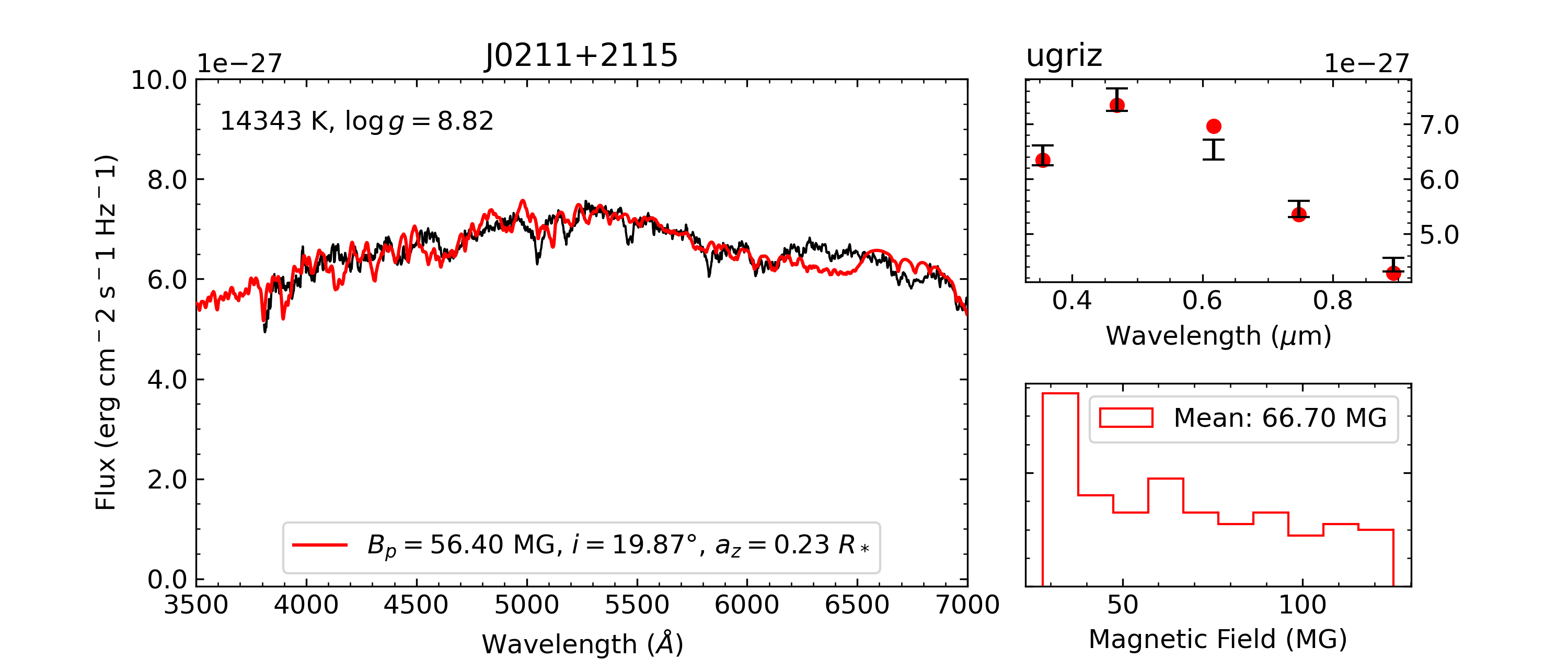}
    \caption{\label{f:magnetic_cant_fit} An example of a magnetic white dwarf which we were unable to find a suitable dipolar geometry.}
\end{figure}

Finally, the last category contains stars that could not be fit within our theoretical framework (38, \autoref{t:nonDA}).
These stars are all probably magnetic white dwarfs of some sort since non-magnetic objects at their fitted effective temperature and surface gravity, as obtained from fitting the $ugriz$ photometry, would show strong hydrogen or helium lines that would be easily recognized.
An example of such a star is presented in \autoref{f:magnetic_cant_fit} where none of the features could be matched using hydrogen-rich models with offset dipole.
The algorithm converged in the parameter space at a place where the surface elements field strength are spread over a large range, effectively washing out spectral features over the entire optical range.
While some of the multitude of features produced this way may match those in the observed spectrum, most of it is probably just random coincidences.
We thus cannot confirm the hydrogen-rich nature of these stars based on our minimum chi-square solution.
It is possible that our failure to find a good solution is due to some large temporal variations of the surface field strength with rotation over timescale much shorter than the integration time of the SDSS spectra.
Combined with that, it is also possible the geometry used here (inclined offset dipole) is too simple an assumption and a much more complex geometry is required.
Alternatively, the main atmospheric composition of those objects might simply not be hydrogen.
In the next paper of that series (Hardy et al. in preparation), we will explore helium-rich compositions and reassess the nature of these objects.


\startlongtable
\begin{deluxetable}{ccccc}
    \tablecaption{\label{t:nonDA} Magnetic non DA stars}
    \tablehead{
        \colhead{} & \colhead{} & \colhead{J name} & \colhead{} & \colhead{}
    }
    \startdata
    J0021+1502 & J0211+0031 & J0211+2115 & J0212+0644 & J0333+0007 \\
    J0732+1642 & J0742+3157 & J0800+0655 & J0822+1201 & J0830+5057 \\
    J0836+1548 & J0847+4842 & J0849+2857 & J0856+2534 & J0922+0504 \\
    J0924+3613 & J0935+4429 & J1121+1039 & J1144+6629 & J1153+1331 \\
    J1202+4034 & J1257+1216 & J1308+8502 & J1348+1100 & J1349+2056 \\
    J1407+3011 & J1453+0652 & J1455+1812 & J1532+1647 & J1623+3546 \\
    J1640+5341 & J1704+3213 & J1724+3234 & J1900+7039 & J2151+0031 \\
    J2247+1456 & J2258+2808 & J2346+3853 &  &  \\
    \enddata
\end{deluxetable}

\startlongtable
\begin{deluxetable}{lccccccc}
    \tablecaption{\label{t:results_good} Model fits with Good dipole fit}
    \tablehead{
        \colhead{J name} & \colhead{$T_{\rm{eff}}$ (K)} & \colhead{$\log{g}$} & \colhead{Mass ($M_\odot$)} & \colhead{$B_p$ (MG)} & \colhead{$i$ (deg)} & \colhead{$a_z$ ($R_*$)} & \colhead{Mean $B$ (MG)}
    }
    \startdata
    J0010+2451 &  10760 (603) &    8.86 (0.49) &   1.129 (0.48) &  11.91 &  55 &  0.30 & 12.25 \\
    J0014$-$1311 &    5725 (10) &  8.17 (0.0065) & 0.696 (0.0086) &  18.77 &  72 &  0.29 & 15.46 \\
    J0114+1607 & 20754 (1979) &       8.95 (0) &   1.178 (0.02) &  27.78 &  84 &  0.10 & 18.42 \\
    J0257+5103 &  20595 (655) &    8.22 (0.12) &   0.753 (0.15) &  12.01 &  56 &  0.30 & 12.17 \\
    J0318+4226 &  10465 (174) &    8.34 (0.17) &   0.817 (0.23) &   4.91 &  24 &  0.15 & 4.89 \\
    J0326+0521 &    30000 (0) &    8.77 (0.76) &   1.099 (0.73) &  17.28 &  63 &  0.26 & 15.29 \\
    J0419$-$0934 &    5952 (59) &   8.34 (0.044) &   0.807 (0.06) &  12.81 &  57 &  0.29 & 12.76 \\
    J0442+1203 &   9529 (337) &   8.38 (0.068) &  0.841 (0.089) &  31.69 &  53 &  0.28 & 32.13 \\
    J0505$-$1722 &    5241 (13) &    7.83 (0.02) &  0.484 (0.022) &   7.42 &  80 & $-$0.17 & 4.53 \\
    J0515+2839 &    6293 (17) &  8.17 (0.0085) &  0.692 (0.011) &   2.22 &  59 &  0.30 & 2.18 \\
    J0556+0521 &    5631 (12) &    8.16 (0.35) &   0.683 (0.44) &  28.99 &  65 &  0.29 & 25.89 \\
    J0601+3726 &  13552 (185) &   8.55 (0.023) &   0.954 (0.03) &   2.30 &  59 &  0.30 & 2.23 \\
    J0604+6413 &  21158 (624) &    8.16 (0.12) &   0.717 (0.11) &  32.08 &  56 &  0.24 & 29.62 \\
    J0648+8403 &  13236 (501) &    8.73 (0.88) &   1.063 (0.89) &   3.70 &  59 &  0.30 & 3.61 \\
    J0649+7521 &    6445 (46) &   8.23 (0.035) &  0.733 (0.047) &  18.24 &  68 &  0.30 & 16.00 \\
    J0651+6242 &    6696 (31) &   8.14 (0.028) &  0.678 (0.035) &   1.22 &  52 &  0.30 & 1.28 \\
    J0725+3214 &  22240 (482) &    8.66 (0.23) &   1.029 (0.44) &  13.83 &  52 &  0.06 & 10.35 \\
    J0732+3646 & 25904 (1239) &    9.31 (0.49) &   1.306 (0.41) & 111.49 &  63 &  0.04 & 78.06 \\
    J0736+4033 &    6320 (30) &    8.29 (0.34) &   0.773 (0.44) &   3.79 &  70 &  0.30 & 3.21 \\
    J0742+2328 &  14852 (573) &    8.76 (0.48) &    1.079 (0.5) &   2.43 &  58 &  0.30 & 2.41 \\
    J0748+3025 & 23078 (1810) &     8.38 (0.5) &   0.863 (0.58) &   8.87 &  55 &  0.28 & 8.81 \\
    J0749+1713 &  21190 (459) &    8.07 (0.16) &   0.664 (0.12) &  13.08 &  55 &  0.19 & 11.37 \\
    J0752+1725 &   9062 (193) &   8.41 (0.069) &  0.861 (0.091) &  12.43 &  56 &  0.30 & 12.52 \\
    J0805+2153 &    30000 (0) &    7.55 (0.25) &   0.444 (0.15) &   3.00 &  17 &  0.30 & 4.08 \\
    J0806+0756 &  15507 (716) &    7.45 (0.25) &   0.360 (0.14) &   2.74 &  69 &  0.19 & 2.13 \\
    J0807+3938 &  15986 (625) &    8.16 (0.25) &   0.713 (0.33) &  37.17 &  56 & $-$0.02 & 24.92 \\
    J0809+3730 &  15993 (898) &    8.51 (0.84) &   0.931 (0.87) &  51.38 &  53 &  0.15 & 42.99 \\
    J0813+2237 &  22398 (720) &    8.42 (0.29) &   0.886 (0.36) &  13.43 &  75 &  0.30 & 10.68 \\
    J0817+2008 &    6989 (82) &   8.30 (0.041) &  0.785 (0.054) &   3.33 &  60 &  0.30 & 3.22 \\
    J0819+3731 &  11576 (458) &    8.32 (0.22) &   0.802 (0.28) &  10.27 &  59 &  0.30 & 10.03 \\
    J0828+2934 & 23855 (1394) &    8.91 (0.14) &   1.163 (0.19) &  35.30 &  67 &  0.23 & 28.94 \\
    J0834+8210 & 28114 (1103) &    9.02 (0.11) &  1.211 (0.092) &  14.59 &  66 &  0.22 & 11.99 \\
    J0839+2000 &  15392 (363) &    8.23 (0.32) &   0.755 (0.39) &   3.01 &  67 &  0.30 & 2.65 \\
    J0840+2712 &  12768 (593) &    8.52 (0.96) &  0.935 (0.097) &   9.85 &  64 &  0.29 & 8.97 \\
    J0841+0223 &    6769 (57) &    8.26 (0.23) &    0.756 (0.3) &   6.50 &  51 &  0.30 & 6.94 \\
    J0842+1539 &  17263 (296) &    7.94 (0.25) &   0.582 (0.24) &  19.66 &  76 & $-$0.29 & 11.28 \\
    J0845+6117 &    5625 (31) &   8.00 (0.061) &  0.587 (0.074) &   0.90 &  23 &  0.08 & 0.79 \\
    J0851+1201 &  11074 (315) &    8.46 (0.11) &   0.893 (0.14) &   2.00 &  78 &  0.30 & 1.54 \\
    J0851+1527 &  14323 (854) &    8.49 (0.18) &   0.916 (0.19) &  29.01 &  67 &  0.27 & 24.88 \\
    J0855+8249 &  23693 (511) &    8.93 (0.34) &   1.173 (0.31) &  12.12 &  68 &  0.26 & 10.07 \\
    J0908+0921 & 29293 (1661) &   8.93 (0.075) &  1.177 (0.072) &  51.55 &  54 &  0.24 & 48.85 \\
    J0911+4202 &  11822 (439) &    8.66 (0.26) &   1.019 (0.31) &  50.08 &  52 &  0.30 & 52.92 \\
    J0914+0544 &  17145 (385) &    8.45 (0.19) &   0.896 (0.24) &   9.42 &  60 &  0.30 & 9.12 \\
    J0925+0113 &   9235 (361) &    8.26 (0.15) &   0.764 (0.19) &   2.66 &  60 &  0.30 & 2.58 \\
    J0931+3219 &  13476 (443) &    8.23 (0.21) &   0.749 (0.21) &   9.04 &  52 &  0.30 & 9.49 \\
    J0933+1022 &   7455 (135) &    7.46 (0.63) &  0.331 (0.044) &   2.18 &  71 &  0.29 & 1.83 \\
    J0934+2945 &  20662 (406) &    8.23 (0.22) &   0.762 (0.23) &  26.92 &  56 &  0.29 & 27.01 \\
    J0934+3927 &   9356 (252) &    8.19 (0.23) &   0.717 (0.28) &   2.10 &  54 &  0.30 & 2.18 \\
    J0934+5033 &   8885 (135) &    8.35 (0.22) &   0.820 (0.28) &   9.12 &  62 &  0.29 & 8.50 \\
    J0942+2052 &  21824 (935) &    8.78 (0.53) &   1.098 (0.53) &  40.55 &  27 & $-$0.01 & 30.22 \\
    J0944+4539 &  16848 (404) &     8.57 (0.2) &    0.975 (0.3) &  16.37 &  87 &  0.20 & 10.92 \\
    J0957+1946 &  14899 (224) &    8.70 (0.23) &   1.047 (0.45) &   4.18 &  82 &  0.26 & 3.00 \\
    J1006+3033 &  10731 (243) &    8.14 (0.56) &   0.689 (0.65) &   4.71 &  50 &  0.30 & 5.10 \\
    J1006+4845 &   9411 (471) &    7.76 (0.25) &    0.469 (0.2) &   8.13 &  49 &  0.27 & 8.44 \\
    J1007+1237 &  18687 (437) &    8.04 (0.24) &   0.646 (0.24) &   6.13 &  71 &  0.30 & 5.15 \\
    J1007+1623 &  11279 (337) &   8.73 (0.058) &  1.061 (0.068) &  13.63 &  42 &  0.30 & 15.89 \\
    J1014+3657 &  10939 (427) &    8.26 (0.41) &   0.763 (0.51) &  14.50 &  74 &  0.28 & 11.47 \\
    J1015+0907 &    7022 (43) &    8.27 (0.19) &   0.760 (0.25) &   4.50 &  47 &  0.30 & 5.00 \\
    J1021$-$1034 &  10485 (124) &   8.59 (0.015) &  0.979 (0.019) &   2.09 &  59 &  0.30 & 2.04 \\
    J1022+2725 &  11581 (606) &    8.70 (0.18) &   1.045 (0.28) &   5.35 &  51 &  0.06 & 4.04 \\
    J1025+6229 &   8578 (311) &    8.52 (0.14) &   0.928 (0.14) &   4.67 &  57 &  0.24 & 4.29 \\
    J1029+1127 &    6691 (72) &     8.20 (0.1) &   0.716 (0.13) &  19.18 &  46 &  0.30 & 21.62 \\
    J1034+0327 &  15756 (406) &   8.80 (0.054) &  1.104 (0.059) &  11.17 &  77 &  0.09 & 7.63 \\
    J1054+5933 &  10704 (383) &    8.30 (0.21) &   0.788 (0.26) &  16.81 &  63 &  0.06 & 11.99 \\
    J1056+6523 &  18833 (616) &    8.18 (0.21) &   0.731 (0.21) &  22.05 &  44 &  0.23 & 22.49 \\
    J1103+0534 &  11233 (369) &     8.67 (0.2) &   1.029 (0.24) &  15.94 &  68 &  0.30 & 13.83 \\
    J1110+6001 &  29215 (361) &    8.92 (0.23) &   1.171 (0.21) &   6.40 &  70 &  0.30 & 5.44 \\
    J1118+0952 &  10438 (325) &    8.44 (0.46) &   0.881 (0.56) &   4.82 &  63 &  0.30 & 4.48 \\
    J1120$-$1150 &  21435 (713) &    9.01 (0.62) &   1.208 (0.56) &   8.57 &  62 &  0.30 & 8.10 \\
    J1122+3223 &  13392 (440) &    8.58 (0.12) &   0.973 (0.12) &  12.22 &  47 &  0.30 & 13.57 \\
    J1123+0956 &   9346 (197) &   8.27 (0.034) &  0.768 (0.045) &   1.85 &  70 &  0.30 & 1.59 \\
    J1128+2649 &  16637 (752) &    7.00 (0.25) &   0.255 (0.11) &   2.34 &  13 & $-$0.12 & 1.48 \\
    J1129+4939 &   9729 (276) &   8.04 (0.085) &    0.625 (0.1) &   5.49 &  54 &  0.30 & 5.68 \\
    J1133+5152 &  21873 (657) &   9.00 (0.042) &  1.201 (0.036) &   6.06 &  65 &  0.30 & 5.54 \\
    J1135+3137 &  16425 (495) &    8.66 (0.18) &   1.028 (0.25) &   2.99 &  47 &  0.19 & 2.79 \\
    J1138$-$0149 &  12534 (435) &    8.28 (0.13) &   0.779 (0.17) &  25.12 &  58 &  0.30 & 24.75 \\
    J1140+6110 &  15976 (627) &    8.11 (0.18) &   0.678 (0.14) &  66.50 &  75 & $-$0.10 & 41.20 \\
    J1144+1717 &  19290 (786) &   8.35 (0.068) &  0.838 (0.088) &   2.00 &  87 &  0.30 & 1.38 \\
    J1148+4827 & 28967 (6860) &       8.97 (0) &  1.194 (0.022) &  34.93 &  65 &  0.29 & 31.25 \\
    J1152+5018 &   9231 (476) &    8.42 (0.34) &   0.869 (0.43) &  13.20 &  50 &  0.30 & 14.34 \\
    J1154+0117 &  29316 (325) &    8.90 (0.21) &    1.162 (0.2) &  35.53 &  87 & $-$0.23 & 22.57 \\
    J1159+6139 &  21591 (887) &     8.99 (0.4) &   1.197 (0.35) &  15.52 &  50 &  0.19 & 14.09 \\
    J1201+6142 &  10412 (482) &   8.52 (0.098) &   0.936 (0.13) &  18.71 &  61 &  0.30 & 17.87 \\
    J1205+3408 &  18753 (424) &    8.16 (0.25) &   0.714 (0.31) &   8.32 &  48 &  0.30 & 9.24 \\
    J1210+2214 &  13177 (426) &   8.48 (0.033) &  0.908 (0.044) &   2.00 &  86 & $-$0.30 & 1.26 \\
    J1217+0828 &  18583 (282) &   8.73 (0.042) &  1.066 (0.047) &   3.50 &  74 &  0.30 & 2.83 \\
    J1222+0015 &  13888 (953) &    9.02 (0.11) &   1.206 (0.14) &  15.47 &  83 &  0.30 & 11.21 \\
    J1222+4811 &   9350 (218) &    8.32 (0.17) &   0.804 (0.22) &   9.29 &  66 &  0.30 & 8.29 \\
    J1223+2307 &  25989 (312) &   8.35 (0.027) &  0.848 (0.035) &   1.61 &  24 &  0.20 & 1.75 \\
    J1224+4155 &  10054 (364) &    8.63 (0.12) &   1.004 (0.15) &  23.99 &  75 &  0.28 & 19.00 \\
    J1227+3855 &  16592 (621) &    8.84 (0.77) &   1.123 (0.76) &   7.07 &  68 &  0.29 & 6.16 \\
    J1227+6612 &  22843 (586) &    8.34 (0.15) &   0.832 (0.19) &  24.67 &  43 &  0.30 & 28.61 \\
    J1234+1248 &    8246 (68) &   7.79 (0.069) &  0.479 (0.071) &   3.41 &  47 &  0.25 & 3.48 \\
    J1248+2942 &    6611 (44) &  7.96 (0.0096) &  0.565 (0.011) &   3.88 &  46 &  0.30 & 4.38 \\
    J1248$-$0229 &  14263 (459) &    8.24 (0.38) &   0.756 (0.47) &   9.07 &  71 &  0.30 & 7.59 \\
    J1250+1549 &   8614 (535) &    7.96 (0.29) &   0.576 (0.32) &  21.41 &  62 &  0.30 & 20.27 \\
    J1254+3710 &  20315 (412) &   8.21 (0.048) &  0.748 (0.061) &   4.11 &  57 &  0.30 & 4.11 \\
    J1254+5612 &  12870 (448) &    8.58 (0.41) &   0.973 (0.48) &  60.43 &  62 &  0.18 & 49.04 \\
    J1257+3414 &  10148 (245) &    8.54 (0.12) &   0.948 (0.15) &  19.68 &  72 &  0.29 & 16.25 \\
    J1300+5904 &    5976 (53) &  7.92 (0.0017) & 0.540 (0.0026) &   4.79 &  35 & $-$0.01 & 3.49 \\
    J1322+5519 &  11637 (305) &    8.50 (0.25) &   0.923 (0.31) &   3.85 &  50 &  0.30 & 4.14 \\
    J1327+1551 &   8619 (871) &    7.79 (0.25) &   0.482 (0.24) &  11.19 &  58 &  0.24 & 10.19 \\
    J1333+6406 &  13159 (659) &   8.84 (0.081) &  1.123 (0.084) &  13.37 &  73 &  0.28 & 10.77 \\
    J1340+6543 &  14638 (537) &    8.17 (0.24) &    0.716 (0.3) &   4.82 &  50 &  0.30 & 5.20 \\
    J1348+3810 &    30000 (0) &   9.13 (0.034) &  1.254 (0.023) &  14.66 &  60 &  0.22 & 12.73 \\
    J1407+4956 &  21372 (585) &    8.70 (0.76) &   1.055 (0.76) &  12.68 &  75 &  0.19 & 9.31 \\
    J1427+3721 &  19876 (743) &   8.92 (0.037) &  1.165 (0.036) &  28.48 &  77 &  0.13 & 19.88 \\
    J1432+4548 &  17262 (526) &    8.23 (0.23) &   0.755 (0.28) &   8.19 &  62 &  0.29 & 7.65 \\
    J1446+5902 &  12169 (418) &    7.93 (0.25) &   0.566 (0.24) &   4.00 &  53 &  0.24 & 3.80 \\
    J1454+4321 &  10647 (608) &   8.33 (0.028) &  0.809 (0.038) &   4.07 &  63 &  0.30 & 3.79 \\
    J1458+2230 &  10823 (347) &     8.69 (0.3) &   1.038 (0.33) &  31.69 &  64 &  0.28 & 28.45 \\
    J1504+0521 & 15383 (1011) &    8.79 (0.44) &   1.101 (0.46) &   4.45 &  54 &  0.30 & 4.60 \\
    J1508+3945 &  18708 (388) &    7.71 (0.58) &  0.473 (0.051) &  20.81 &  56 &  0.14 & 16.94 \\
    J1514+0744 &  10572 (727) &    8.27 (0.09) &   0.770 (0.12) &  37.89 &  53 &  0.30 & 39.69 \\
    J1514+1520 &   9252 (293) &    8.52 (0.11) &   0.931 (0.14) &   2.32 &  53 &  0.30 & 2.43 \\
    J1515+2445 &   8718 (187) &    8.41 (0.29) &   0.862 (0.37) &  14.45 &  62 &  0.29 & 13.54 \\
    J1524+1856 &  14611 (553) &    8.75 (0.12) &   1.077 (0.13) &  12.75 &  61 &  0.26 & 11.62 \\
    J1538+0842 &   9656 (221) &   8.60 (0.006) &  0.983 (0.008) &  11.52 &  49 &  0.29 & 12.31 \\
    J1538+5306 &  13965 (451) &    8.33 (0.73) &   0.818 (0.82) &  10.97 &  36 &  0.01 & 8.17 \\
    J1542+0348 &   9835 (318) &    8.61 (0.26) &   0.987 (0.32) &   5.41 &  34 &  0.28 & 6.55 \\
    J1543+3432 &  29554 (563) &    8.88 (0.24) &   1.151 (0.23) &   2.65 &  60 &  0.30 & 2.56 \\
    J1548+2451 &  21161 (344) &   8.99 (0.038) &  1.198 (0.033) &   7.29 &  73 &  0.30 & 6.01 \\
    J1647+3709 &  16762 (296) &   8.42 (0.089) &   0.878 (0.11) &   2.11 &  52 &  0.30 & 2.24 \\
    J1650+3411 &   9797 (299) &   8.45 (0.079) &    0.884 (0.1) &   4.36 &  55 &  0.30 & 4.44 \\
    J1652+3334 &   8965 (178) &   8.15 (0.061) &  0.687 (0.078) &   4.91 &  49 &  0.30 & 5.38 \\
    J1652+3528 &  11615 (501) &    8.13 (0.71) &  0.683 (0.078) &  10.13 &  62 &  0.30 & 9.56 \\
    J1659+4401 &  28957 (811) &   9.22 (0.098) &  1.283 (0.084) &   3.90 &  50 &  0.30 & 4.20 \\
    J1707+3532 &  22872 (502) &   8.91 (0.046) &  1.161 (0.044) &   2.13 &  62 &  0.30 & 2.01 \\
    J1714+3918 &    6577 (26) &   7.77 (0.038) &  0.461 (0.038) &   2.40 &  50 &  0.30 & 2.58 \\
    J1717+2620 & 28729 (1290) &    9.15 (0.39) &   1.259 (0.31) &  21.26 &  66 &  0.30 & 19.14 \\
    J1720+5612 & 24352 (2526) &    8.76 (0.15) &   1.086 (0.23) &  19.03 &  60 &  0.10 & 14.34 \\
    J1723+5407 &  11270 (315) &   8.90 (0.049) &  1.151 (0.047) &  39.18 &  58 &  0.09 & 29.50 \\
    J1729+5632 &  10950 (471) &    8.75 (0.12) &   1.074 (0.13) &  28.57 &  67 &  0.22 & 23.34 \\
    J2025+1310 &  17686 (622) &    8.14 (0.12) &   0.701 (0.11) &  10.62 &  62 &  0.30 & 10.00 \\
    J2046$-$0710 &    8350 (74) &   8.25 (0.084) &   0.755 (0.11) &   2.41 &  64 &  0.30 & 2.22 \\
    J2052$-$0016 &  20823 (428) &    8.87 (0.41) &   1.143 (0.39) &  14.00 &  78 &  0.14 & 9.76 \\
    J2149$-$0728 &  22642 (750) &    8.37 (0.14) &   0.852 (0.18) &  45.09 &  66 &  0.17 & 34.93 \\
    J2227+1753 &    6548 (28) &   7.93 (0.006) & 0.550 (0.0071) &   1.30 &  40 &  0.29 & 1.53 \\
    J2332+2658 &   9497 (197) &    8.43 (0.15) &    0.870 (0.2) &   2.57 &  57 &  0.30 & 2.57 \\
    J2348+2535 &  25122 (396) &    8.53 (0.47) &   0.959 (0.54) &   6.30 &  44 &  0.30 & 7.26 \\
    \enddata
\end{deluxetable}


\startlongtable
\begin{deluxetable}{lccccccc}
    \tablecaption{\label{t:results_bad} Model fits with Bad dipole fit}
    \tablehead{
        \colhead{J name} & \colhead{$T_{\rm{eff}}$ (K)} & \colhead{$\log{g}$} & \colhead{Mass ($M_\odot$)} & \colhead{$B_p$ (MG)} & \colhead{$i$ (deg)} & \colhead{$a_z$ ($R_*$)} & \colhead{Mean $B$ (MG)}
    }
    \startdata
    J0006+0755 &   8939 (189) &    8.26 (0.11) &   0.759 (0.14) &  13.70 &  48 &  0.30 & 15.05 \\
    J0234+2648 &  15103 (703) &    8.39 (0.31) &   0.853 (0.39) &  45.27 &  51 &  0.19 & 40.78 \\
    J0304$-$0025 &  13190 (533) &    8.32 (0.11) &   0.807 (0.15) &  11.49 &  59 &  0.24 & 10.42 \\
    J0331+0045 &  16796 (268) &   8.57 (0.076) &  0.974 (0.093) &  12.82 &  58 &  0.29 & 12.52 \\
    J0345+0034 &    7345 (77) &    7.99 (0.25) &    0.589 (0.3) &   2.43 &  45 &  0.30 & 2.75 \\
    J0537+6759 &    7661 (45) &   8.33 (0.028) &  0.804 (0.037) &   1.13 &  61 &  0.30 & 1.09 \\
    J0632+5559 &    9876 (37) &  8.52 (0.0086) &  0.930 (0.011) &   0.81 &  45 &  0.30 & 0.92 \\
    J0758+3544 & 22693 (1052) &    8.96 (0.12) &   1.183 (0.11) &  30.90 &  36 &  0.30 & 38.00 \\
    J0803+1229 &  20441 (505) &  8.93 (0.0081) & 1.172 (0.0081) &  37.43 &  31 &  0.26 & 44.25 \\
    J0804+1827 &  11731 (573) &   8.66 (0.012) &  1.019 (0.016) &  72.44 &  64 &  0.30 & 66.52 \\
    J0816+0412 &  12735 (433) &    7.84 (0.25) &   0.517 (0.22) &  10.33 &  65 &  0.26 & 9.02 \\
    J0849+0037 &  15424 (453) &    8.33 (0.18) &   0.818 (0.22) &  17.93 &  57 &  0.20 & 15.52 \\
    J0855+1640 &  16332 (709) &    8.43 (0.63) &   0.885 (0.72) &  15.34 &  34 &  0.30 & 19.11 \\
    J0858+4126 &    7013 (84) &   8.37 (0.057) &  0.829 (0.076) &   2.64 &  41 &  0.30 & 3.09 \\
    J0907+3538 & 18101 (1029) &    8.76 (0.14) &   1.085 (0.17) &  16.44 &  49 &  0.26 & 16.92 \\
    J0930$-$0126 &  14944 (823) &    8.91 (0.39) &   1.157 (0.37) &  21.66 &  50 &  0.30 & 23.41 \\
    J0937+1021 & 19955 (1174) &    8.98 (0.64) &    1.193 (0.6) & 288.08 &  66 &  0.14 & 217.85 \\
    J1018+0111 &  12845 (753) &    8.70 (0.14) &   1.045 (0.16) &  76.07 &  16 &  0.05 & 64.24 \\
    J1018+3033 &  21298 (437) &    8.78 (0.15) &   1.098 (0.19) &  50.92 &  78 &  0.24 & 37.68 \\
    J1020+3626 & 19101 (1071) &    8.31 (0.18) &   0.813 (0.23) &  59.79 &  69 &  0.15 & 44.69 \\
    J1022+1949 &   8673 (189) &    8.38 (0.33) &   0.838 (0.42) &   2.65 &  30 &  0.30 & 3.37 \\
    J1033+2309 & 26025 (1171) &    8.77 (0.15) &   1.097 (0.15) & 230.49 &  40 &  0.26 & 254.52 \\
    J1035+2126 &    6882 (53) &   8.09 (0.021) &  0.644 (0.027) &   2.63 &  48 &  0.30 & 2.89 \\
    J1126+0906 &  11623 (496) &    8.86 (0.04) &   1.132 (0.04) &  26.08 &   2 &  0.28 & 35.13 \\
    J1130+3057 &  10767 (384) &    8.55 (0.18) &   0.952 (0.22) &   3.94 &  30 &  0.29 & 4.96 \\
    J1206+0813 &  15163 (998) &    8.80 (0.26) &   1.103 (0.27) & 180.54 &  50 &  0.30 & 194.44 \\
    J1216$-$0026 & 16409 (1263) &    8.83 (0.18) &   1.118 (0.27) &  63.24 &  53 &  0.27 & 63.85 \\
    J1242+4548 &  11667 (393) &    8.56 (0.22) &   0.963 (0.39) &  10.93 &  21 &  0.30 & 14.57 \\
    J1351+5419 & 13937 (1038) &    8.43 (0.27) &   0.882 (0.34) & 368.52 &  34 &  0.07 & 303.55 \\
    J1508+2150 &  19995 (406) &    8.17 (0.12) &   0.726 (0.15) &  23.78 &  42 &  0.30 & 27.78 \\
    J1511+4220 &  11595 (411) &    8.72 (0.09) &   1.054 (0.11) &  14.01 &  40 &  0.30 & 16.61 \\
    J1517+6105 &  10751 (782) &    8.83 (0.12) &   1.118 (0.14) &  10.27 &  47 &  0.11 & 8.45 \\
    J1535+4213 & 18696 (2151) &    7.70 (0.25) &   0.469 (0.18) &   5.99 &  83 & $-$0.29 & 3.65 \\
    J1557+0411 & 24847 (1220) &   9.04 (0.044) &  1.218 (0.037) &  43.83 &  55 &  0.25 & 41.65 \\
    J1601+0442 &  20251 (983) &    8.52 (0.44) &   0.942 (0.52) &  92.59 &  18 &  0.03 & 75.75 \\
    J1603+1409 &  10547 (348) &   8.62 (0.017) &  0.996 (0.022) &  48.88 &  47 &  0.23 & 48.74 \\
    J1604+4908 &  10717 (478) &    8.67 (0.15) &   1.023 (0.18) &  67.10 &  40 &  0.30 & 80.07 \\
    J1608+0644 &   9407 (256) &    8.51 (0.41) &    0.925 (0.5) &  22.71 &  37 &  0.30 & 27.63 \\
    J1633+1942 &  16393 (230) &   8.13 (0.045) &  0.695 (0.056) &   3.61 &  54 &  0.30 & 3.72 \\
    J1648+4618 &  14608 (473) &    8.31 (0.11) &   0.803 (0.14) &  19.48 &  50 &  0.30 & 21.18 \\
    J1709+2341 &  13784 (817) &    8.53 (0.19) &   0.943 (0.26) &  12.07 &  50 &  0.28 & 12.72 \\
    J2117$-$0736 & 26033 (1002) &    9.02 (0.12) &   1.211 (0.15) &   6.24 &  51 &  0.30 & 6.71 \\
    J2149+0048 &  14121 (484) &    8.67 (0.17) &   1.032 (0.23) &   9.57 &  54 &  0.27 & 9.55 \\
    J2218$-$0000 &  16678 (702) &    8.84 (0.12) &   1.127 (0.12) &  96.00 &  20 &  0.28 & 123.63 \\
    J2322+0039 &  20241 (573) &    8.97 (0.58) &   1.188 (0.52) &  11.00 &  19 &  0.10 & 10.12 \\
    \enddata
\end{deluxetable}

\section{Discussion} \label{s:discussion}

\begin{figure}
    \centering
    \includegraphics[width=0.7\linewidth]{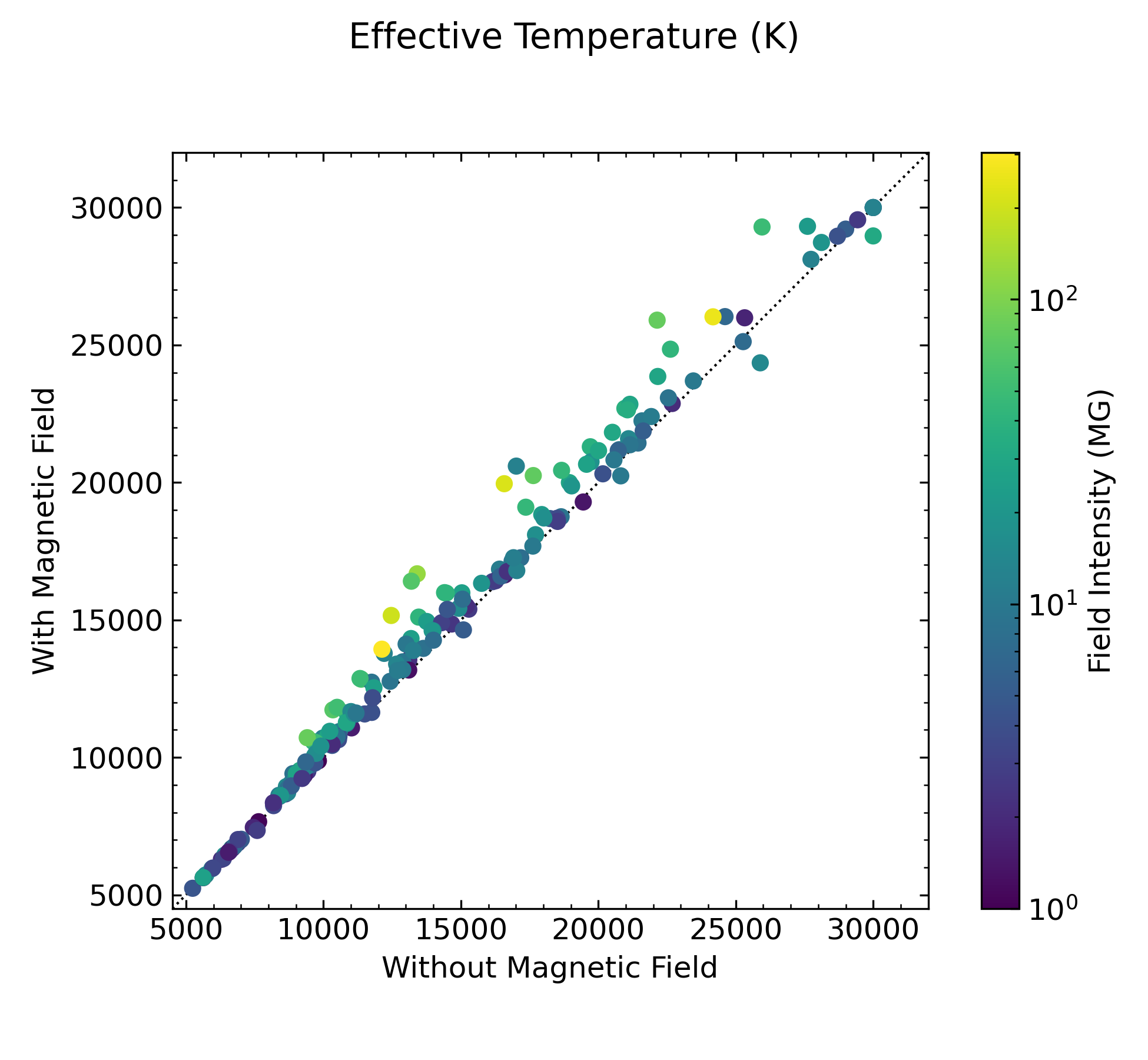}
    \caption{\label{f:compare_teff} Comparison of effective temperatures obtained from non-magnetic vs. magnetic synthetic spectra. The color scale indicate the strength of the magnetic field.}
\end{figure}

\begin{figure}
    \centering
    \includegraphics[width=0.7\linewidth]{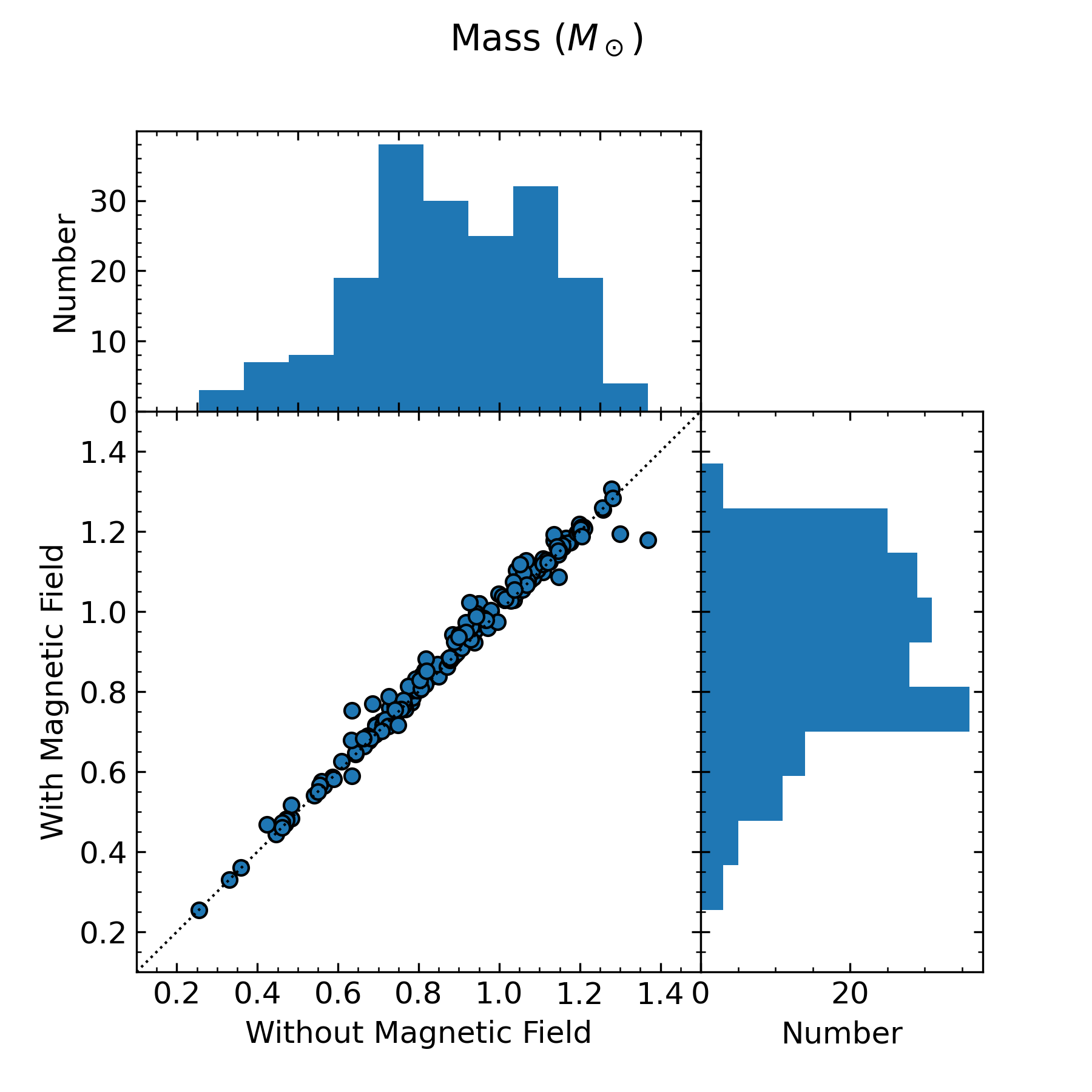}
    \caption{\label{f:compare_mass} Comparison of stellar masses obtained from non-magnetic vs. magnetic synthetic spectra. The corresponding mass distribution histograms are also shown on the right and upper part of the figure. }
\end{figure}

From our original sample, we are left with 185 objects that we could fit with various degrees of quality using a dipolar geometry.
\autoref{f:compare_teff} shows that including the effect of the magnetic field in the synthetic color calculations tends to systematically increase the determined effective temperature compared to that obtained from non-magnetic models (the difference can be as high as 3700K for stars with the largest field intensity).
As explained in \autoref{ss:photo}, this is a direct consequence of the shifting of absorption lines across adjacent photometric bands as the magnetic field strength increases.
To compensate for this increase in effective temperature, it becomes necessary to decrease the stellar radii in order to reproduce the photometric data.
Hence, the derived masses are also increased by up to \Msol{0.1} compared to masses determined from non-magnetic models (see \autoref{f:compare_mass}).

It is interesting to discuss our sample in light of the results drawn from the 40 pc sample recently presented by \citet{Bagnulo2022}.
According to their interpretation, MWDs form two distinct populations distinguished by their very different typical masses.
The most massive magnetic white dwarfs, according to their hypothesis, would mostly be the result of binary mergers, producing strong magnetic fields that would be present at the surface as soon as the star enters the cooling sequence.
Such mergers are expected to be rapidly rotating.
On the other hand, young MWDs with mass below about $\sim$ \Msol{0.75} appear to be extremely rare and magnetism would gradually appear only after 2 Gyr or 3 Gyr, possibly due to the relaxation of a preexisting field buried in the interior or generated by a dynamo mechanism triggered by the onset of core crystallization \citep{Isern2017, Schreiber2021}.

\begin{figure}
    \centering
    \includegraphics[width=0.7\linewidth]{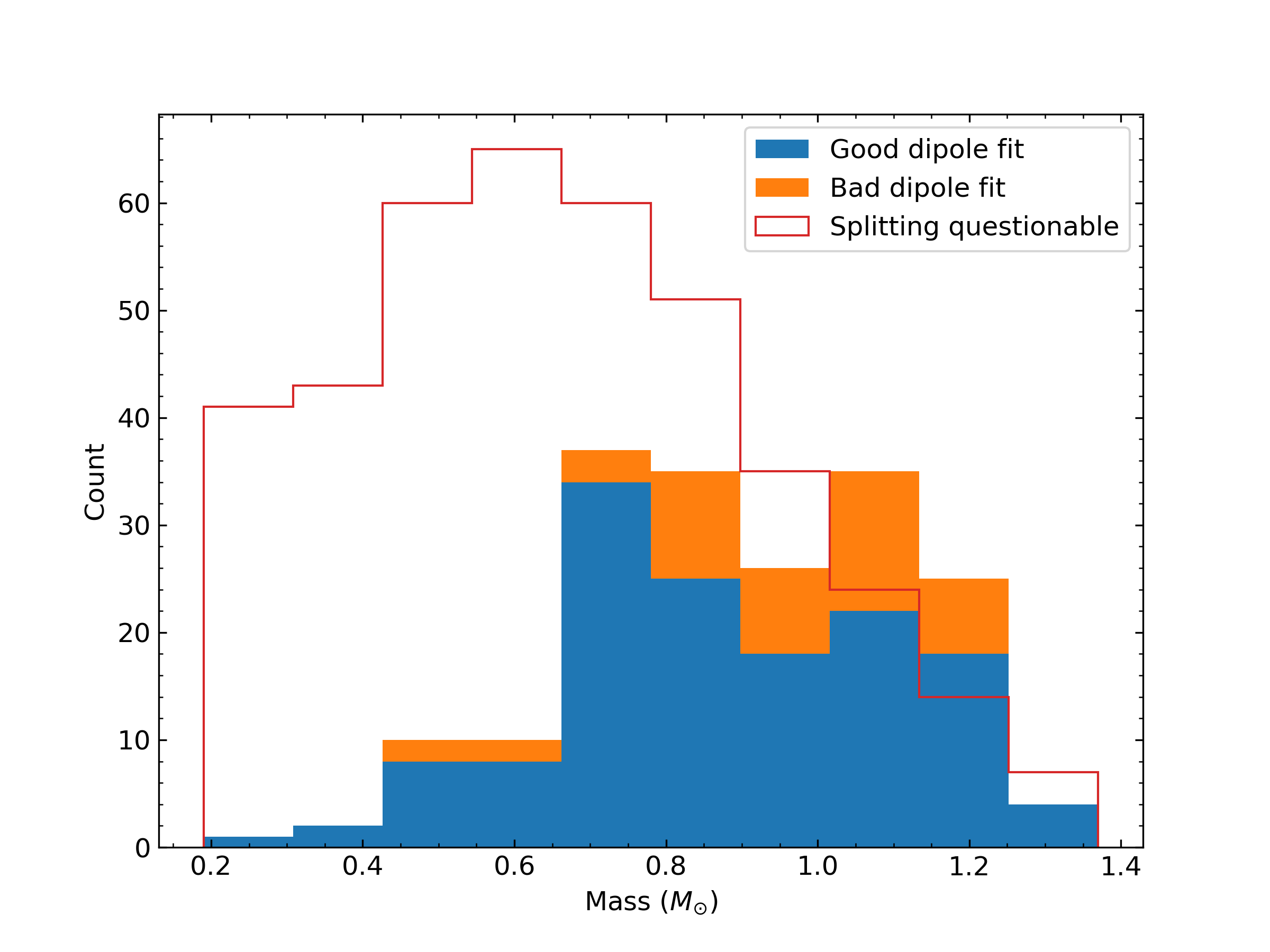}
    \caption{\label{f:mass_dist} Mass distribution for the stars in our sample.}
\end{figure}

\autoref{f:mass_dist} takes a closer look at the mass distribution of our sample as well as all the stars that we rejected in the previous section.
First, we see that the stars previously classified DAH that we dismissed as magnetic form a relatively broad distribution centered on $\sim$\Msol{0.6}, as expected for a population of non-magnetic white dwarfs.
This unusually large distribution can easily be explained by the fact that those stars are almost twice as distant, on average, than the stars that we successfully fit with a dipole.
These were thus objects with the very lowest signal-to-noise ratio spectra of the sample, and the observed splitting that led to their classification as DAH was most probably just noise.
Being more distant, the uncertainties on the photometric data and parallax measurements were also much larger, explaining the extremely broad mass distribution.

\begin{figure}
    \centering
    \includegraphics[width=0.7\linewidth]{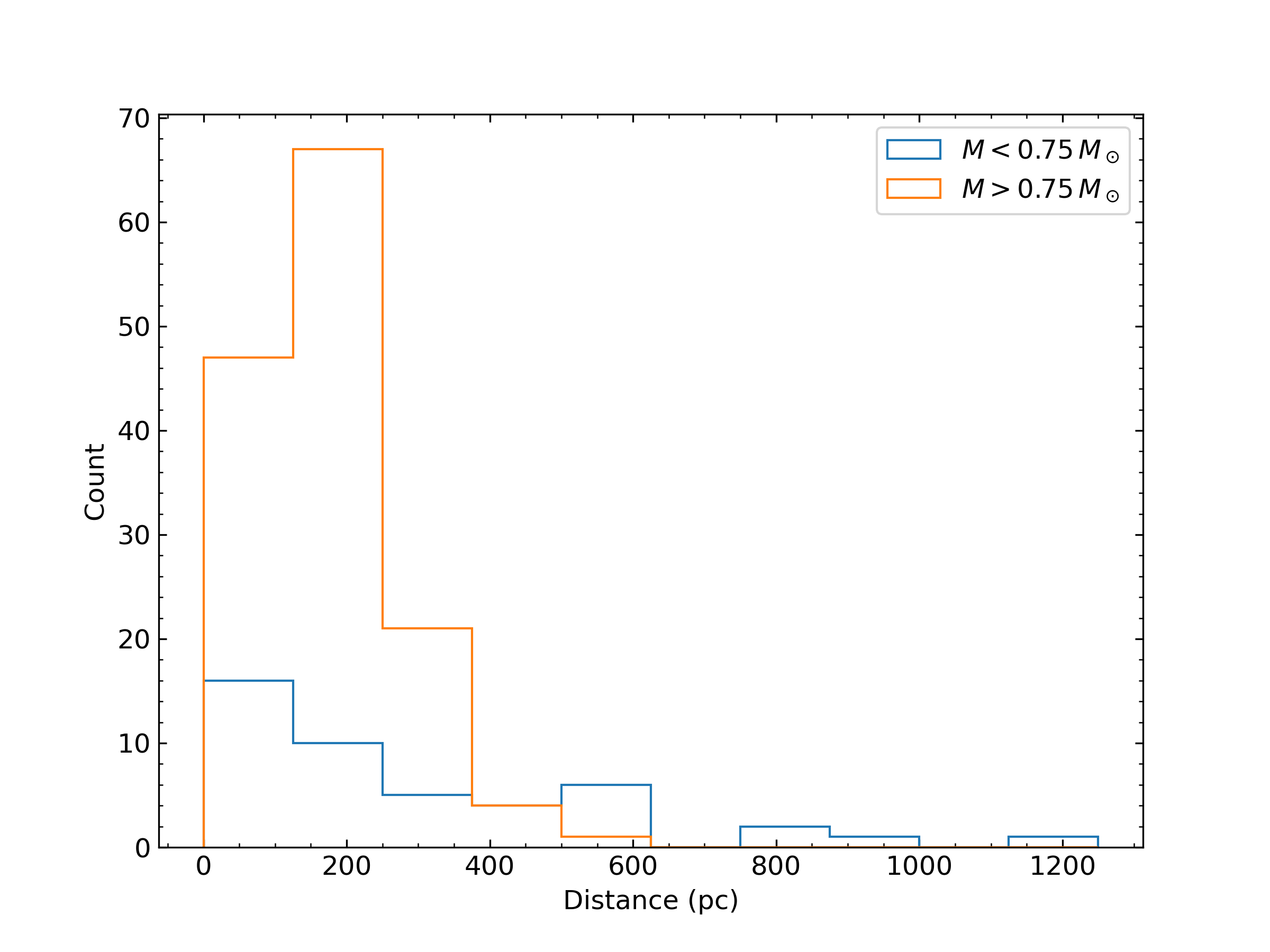}
    \caption{\label{f:distance_mag_dist} Distance distribution of confirmed Magnetic DA stars, divided into two groups: One with masses below \Msol{0.75}, the other with masses above.}
\end{figure}

\begin{figure}
    \centering
    \includegraphics[width=0.7\linewidth]{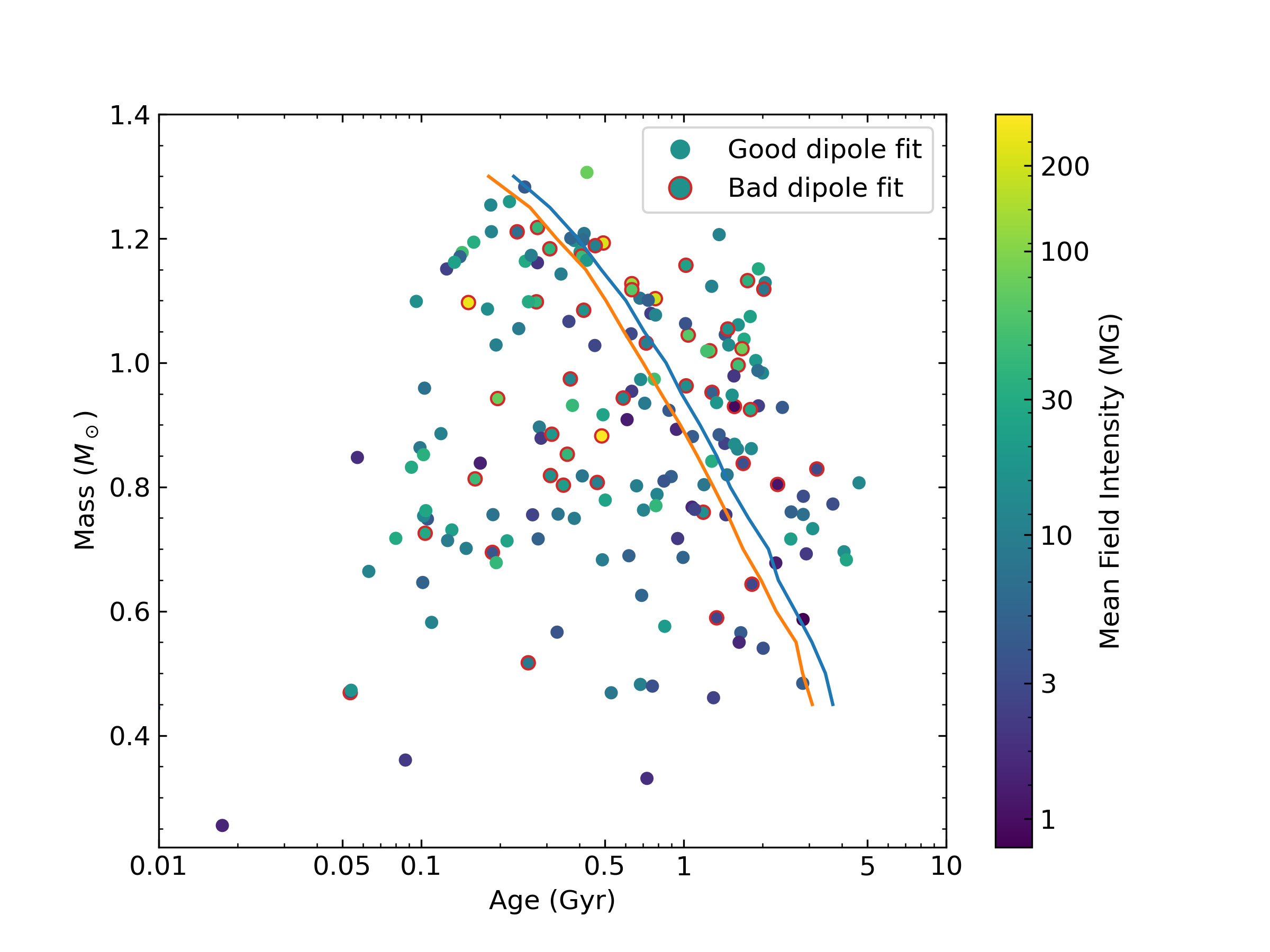}
    \caption{\label{f:mass_age} Stellar mass vs. age, color coded with the mean surface magnetic field for the confirmed stars in our sample. The blue and orange lines represent the onset of crystallization for stars with a thick ($q_{He}=10^{-2}$, $q_H=10^{-4}$) and thin ($q_{He}=10^{-2}$, $q_H=10^{-10}$) hydrogen layer, respectively.}
\end{figure}

Next, we observe a sudden increase in the number of MWDs at a mass of about \Msol{0.7}.
This may indicate that the most massive MWDs have a different origin than the ones at lower mass, as suggested by \citet{Bagnulo2022}.
We must, however, remain cautious with this interpretation as our sample is mostly drawn from a magnitude-limited survey that certainly suffer from a number of selection biases difficult to quantify.
Nevertheless, a look at the distance distribution of our sample in \autoref{f:distance_mag_dist} suggests that the preponderance of massive white dwarfs in our magnetic sample is real and not merely an artifact of various selection effects.
Indeed, while it can be argued that massive white dwarfs may be, due to their much smaller radii, underrepresented in a magnitude limited sample, the fact that we see a cutoff in the number of magnetic stars with mass above \Msol{0.75} starting at about 250 pc suggest that we are perfectly capable of identifying such objects in SDSS at smaller distances.
Meanwhile, MWDs less massive than \Msol{0.75} are much larger (and thus brighter) and would be, at similar effective temperature, easily identify in that same volume, assuming that whatever selection bias affecting our sample is the same at all white dwarf masses.
According to the reasoning of \citet{Bagnulo2022} described above, this is totally to be expected if magnetism in lower mass objects only appears much later on the cooling sequence.
The big difference with our sample is that contrary to what is observed in the 40 pc sample, low mass MWDs are not completely absent for ages younger than 1-2 Gyr (see \autoref{f:mass_age}).
Taking only objects within 250 pc does not change the overall picture emerging from \autoref{f:mass_age} much; young MWDs less massive than \Msol{0.75}, while much rarer than heavier ones, do exist.
Their magnetic field strength, however, never reach values above $\sim$30 MG as is regularly found at higher masses.
It is possible, however, that whatever mechanism is responsible for the presence of strong magnetic fields in young low mass white dwarfs is sufficiently rare for it to not be statistically present in the 40 pc sample and that only looking at a much larger volume does this additional channel manifest its presence.
To summarize, our sample does suggest white dwarfs with strong magnetic field and mass above \Msol{0.75} are much more numerous than their lower-mass counterpart.
While selection biases may play a role in the exact proportion, this assessment is unlikely to change with a large volume-limited sample.
The young low mass MWDs in our sample could represent a separate channel not observed in the 40 pc sample that is superposed to the one where magnetism only start to show up after the onset of crystallization in the core.

\begin{figure}
    \centering
    \includegraphics[width=0.7\linewidth]{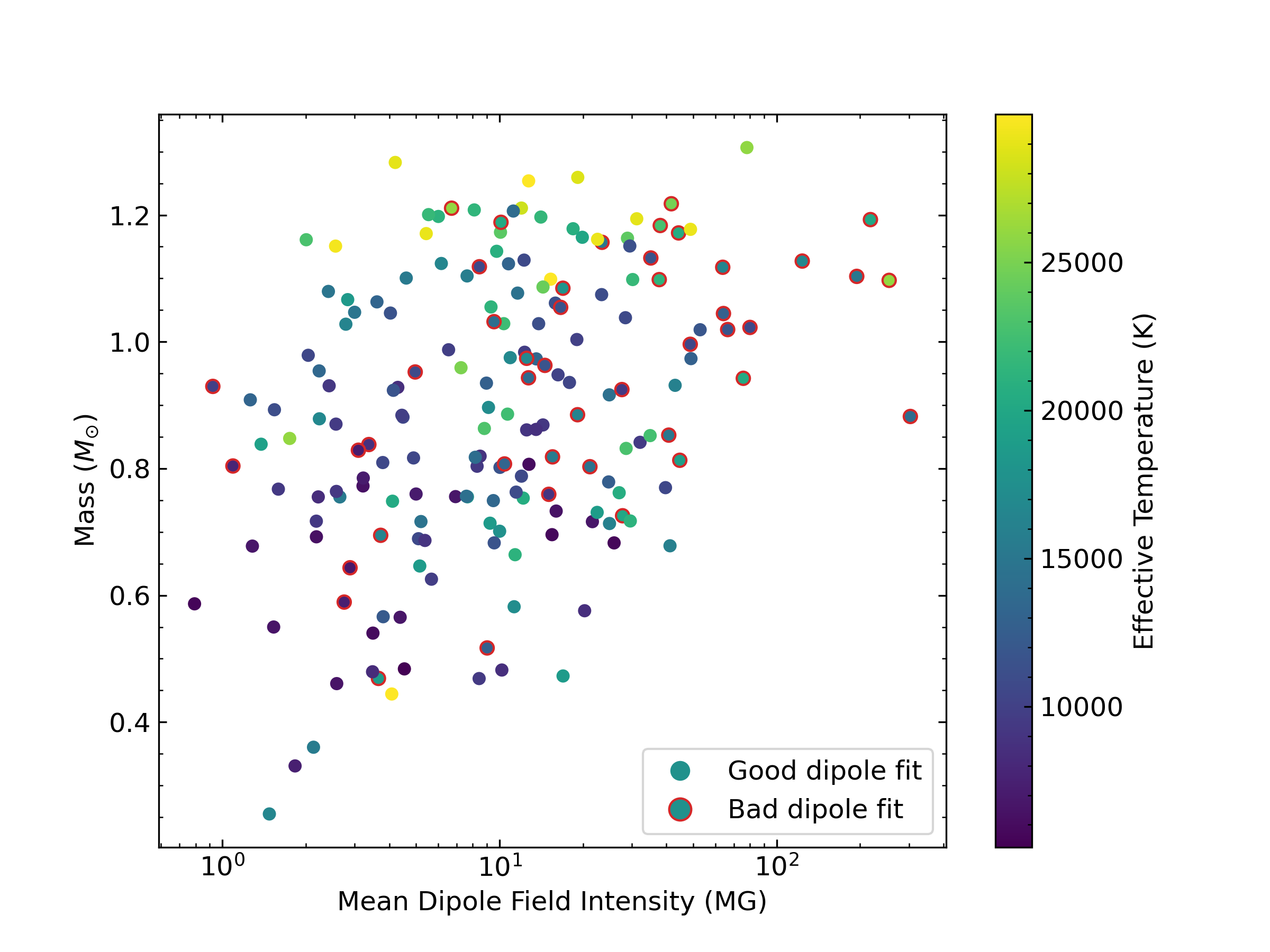}
    \caption{\label{f:mass_field} Mass vs. magnetic field intensity for the confirmed magnetic stars in our sample. The color scale shows effective temperature for each star.}
\end{figure}

\begin{figure}
    \centering
    \includegraphics[width=0.7\linewidth]{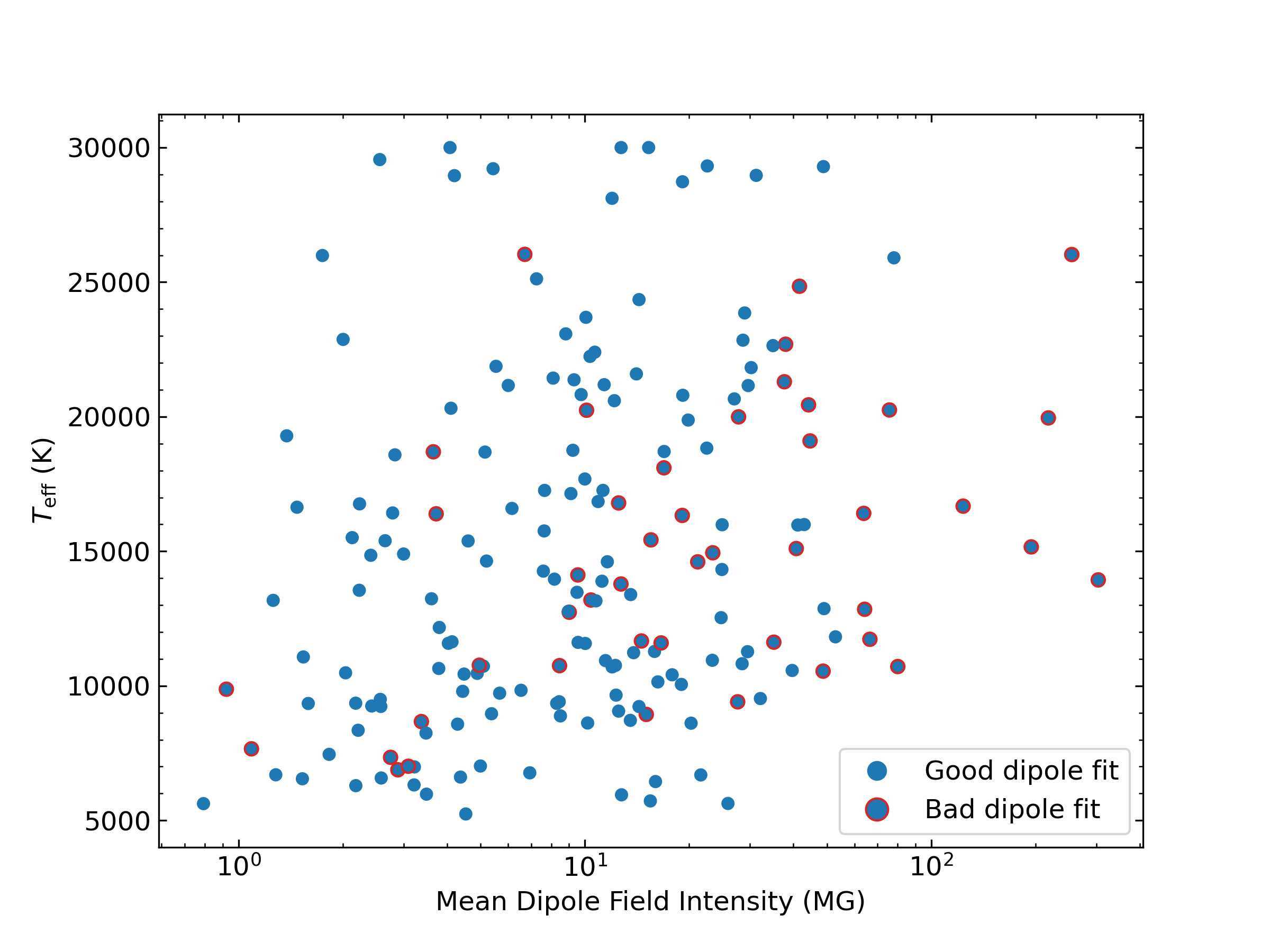}
    \caption{\label{f:teff_field} Effective temperature vs. magnetic field intensity for the confirmed magnetic stars in our sample.}
\end{figure}

Are there other signatures for different channels/origin that emerge from the homogeneous analysis of the sample?
\autoref{f:mass_field} shows little correlation between the mean dipole field intensity (mean field intensity on the visible surface of the star) and stellar mass, except for the fact that all the largest fields are found only in high mass white dwarfs.
The stars that were badly fit with a dipole are slightly more prevalent among the most massive ($M \ge$ \Msol{0.75}) and most magnetic ($B \ge$ 10 MG) but can nevertheless be found at any mass, field or effective temperature.
We can see a slight tendency for heavier stars to be hotter, as there is a higher density of $> 25000$ K stars near \Msol{1.2}, although this could be a selection bias since more massive stars are smaller, and need to be hotter and more luminous to be seen at a given magnitude.
Looking instead at the effective temperature vs. the mean dipole field intensity (\autoref{f:teff_field}) does not highlight any discernable pattern that would allow us to distinguish between different groups that would possibly have different origins.
Such distinction may require a detailed analysis based on time-resolved spectroscopy in order to pinpoint possible differences in field structures.

\section{Conclusion} \label{s:conclu}

In this study, we used state-of-the-art magnetic synthetic models to perform the first large homogeneous analysis of hydrogen-rich MWDs since \citeyear{Kulebi2009}.
We carefully examined 651 white dwarfs that were labeled as magnetic or DAH compiled in the Montreal White Dwarf Database.
Since most of the sample is drawn from the Sloan Digital Sky Survey, only white dwarfs with magnetic fields large enough to produce an undeniable line splitting (above $\sim$1-2 MG) could be assessed.
From our original sample, 400 stars are reclassified as non-magnetic white dwarfs as the alleged splitting reported in the literature did not hold up against attentive scrutiny.
We find that 140 stars could be relatively well fitted with an offset dipole while another 45 were clearly DAH but a dipolar fit was passable.
We were also unable to fit 38 objects with our theoretical framework, suggesting that these objects either have a much more complex field structure or are simply not hydrogen-rich.
Finally, the depth of the absorption line for 28 stars were too shallow compared to model predictions, possibly due to non-homogeneous/mixed surface compositions.

We find a sharp increase in the number of MWDs starting at a mass of about \Msol{0.70}.
The strongest magnetic fields are also found only among the most massive white dwarfs (no field above 30 MG are found for $M \leq$ \Msol{0.70}).
Stars that are well and badly fitted with a dipole geometry appear to be disseminated evenly across the whole range of masses, effective temperatures and magnetic field strengths, leaving little clues about their possibly different origins.
The stars badly fitted with a dipole could have a much more complex field structure or simply vary on short time scaled compared to the spectra integration time.
Signature of the evolutionary origin of the magnetic field, if present, will have to wait for the availability of time-resolved spectroscopy in order to better quantify the details of their field structure.

A study of the 40 pc sample by \citet{Bagnulo2022} recently suggested that magnetism in white dwarf star with masses below \Msol{0.75} only start to appear after about 2 Gyr, possibly as the result of a dynamo mechanism that is triggered by the start of core crystallization.
However, based on our sample, it appears that there is an additional channel that is capable of producing very young magnetic white dwarfs with $M \leq$ \Msol{0.70}.
With about two dozen such object found in a volume much larger than the 40 pc sample, it may, however, not be statistically significant to not find any in the local sample.
Because our sample certainly suffer from numerous selection biases, we refrain from quantifying the proportion of such objects as a function of age or mass.

To conclude, detailed analysis of large samples of magnetic white dwarfs are needed to provide new insights on the nature and evolution of such stars.
Our study, while inconclusive, represent the largest effort in that endeavor since the work of \citet{Kulebi2009}, almost doubling the number of magnetic white dwarfs meticulously analyzed in a homogeneous fashion.
Our investigation established the existence of various families of magnetic objects that deserve further scrutiny, especially those not fit well with a dipole geometry or those with shallow absorption features.
Some object that may not have a hydrogen-rich surface composition will be examined in detail in the next paper of this series.
Time-resolved spectroscopy, spectropolarimetry and much higher signal-to-noise/resolution observations are highly awaited for advancing our understanding of the origin and evolution of magnetic white dwarfs.
The unique models and fitting tool developed for this project have certainly proven themselves as reliable, and will be the basis of studies that will push even further our understanding of MWDs.

\section*{Data Availability}

The data underlying this article are available on the Montreal White Dwarf Database\footnote{\url{https://www.montrealwhitedwarfdatabase.org/}} \citep{MWDD}.

\bibliographystyle{apj}
\bibliography{references}

\appendix


\startlongtable
\begin{deluxetable}{cccrr}
    \tablecaption{Object Names and Coordinates\label{t:names}}
    \tablehead{
        \colhead{J Name} & \colhead{Gaia Source ID} & \colhead{MWDD ID} & \colhead{R.A.} & \colhead{Decl.}
    }
    \startdata
    J0002$-$4310 & 4994877094997259264 & LAWD 96                        & 0.54831 & $-$43.16834 \\
    J0003$-$0153 & 2448780654695945728 & SDSS J000321.59$-$015310.8       & 0.84005 & $-$1.88640 \\
    J0005+1934 & 2798285095877328512 & SDSS J000528.70+193457.4       & 1.36971 & 19.58270 \\
    J0006+0755 & 2746037712074342784 & SDSS J000603.80+075514.4       & 1.51576 & 7.92059 \\
    J0007+2407 & 2849716401534278784 & SDSS J000701.59+240744.7       & 1.75682 & 24.12903 \\
    J0010+2429 & 2849087274724833408 & SDSS J001050.27+242946.6       & 2.70947 & 24.49616 \\
    J0010+2451 & 2850233721754720896 & SDSS J001034.95+245131.2       & 2.64549 & 24.85856 \\
    J0012+5025 & 395234439752169344 & GJ 1004                        & 3.05832 & 50.42005 \\
\enddata
\end{deluxetable}

\newcounter{pagecounter}
\newcounter{allfitspages}
\setcounter{allfitspages}{62}

\makeatletter
\@whilenum{\value{pagecounter} < \value{allfitspages}}\do{%
    \stepcounter{pagecounter}
    \centering{
        \includegraphics[page=\value{pagecounter}, width=0.94\linewidth]{figures/fits/all_fits}
        \ifnum \value{pagecounter} = 1 {
                    \phantomsection
                    \label{f:all_fits}
                } \fi
        \clearpage
    }
}
\makeatother



\end{document}